# Weaving Equity into Infrastructure Resilience Research and Practice: A Decadal Review and Future Directions

**Authors:** Natalie Coleman [1*], Xiangpeng Li [2], Tina Comes [3], Ali Mostafavi [4]

**Affiliations:**
[1] Ph.D. Student, Zachry Department of Civil and Environmental Engineering, Urban Resilience.AI Lab, Texas A&M University, College Station, TX, United States of America; email: ncoleman@tamu.edu
*Corresponding author
[2] Ph.D. Student, Zachry Department of Civil and Environmental Engineering, Urban Resilience.AI Lab, Texas A&M University, College Station, TX, United States of America; email: xplli@tamu.edu
[3]Full Professor and Director, TPM Resilience Lab, TU Delft, Delft, South Holland, Netherlands; email: T.Comes@tudelft.nl
[4] Associate Professor, Zachry Department of Civil and Environmental Engineering, Urban Resilience.AI Lab, Texas A&M University, College Station, TX, United States of America; email: amostafavi@civil.tamu.edu

## Abstract
Disasters amplify existing inequalities, and infrastructures play a crucial role in this process. They determine the access of vulnerable communities to clean water, food, healthcare, and electricity. Increasingly, the need to account for equity in infrastructure resilience has been recognized. After about a decade of research in this domain, what is missing is a systematic overview of the state-of-the art and a research agenda across different infrastructures and hazards. It is now imperative to evaluate the current progress and gaps. This paper presents a systematic review of equity literature on disrupted infrastructure during a natural hazard event. Following a systematic review protocol, we collected, screened, and evaluated almost 3,000 studies. Our analysis focuses on the intersection within the dimensions of the eight-dimensional assessment framework that distinguishes *focus of the study* ; the *methodological approaches* and the *equity dimensions* (distributional-demographic, distributional-spatial, procedural, and capacity equity). To conceptualize the intersection of the different dimensions of equity and how they are applied to different contexts, we refer to "pathways", which identify how equity is constructed, analyzed, and used. Significant findings show that (1) the interest in equity in infrastructure resilience has exponentially increased, (2) the majority of studies are in the US and by extension in the global north, (3) most data collection use descriptive and open-data and none of the international studies use location-intelligence data. The most prominent equity conceptualization is distributional equity, such as the disproportionate impacts to vulnerable populations and spaces. The most common pathways to study equity connect distributional equity to the infrastructure's power, water, and transportation in response to flooding and hurricane storms. Other equity concepts or pathways, such as connections of equity to decision-making and building household capacity, remain understudied. Future research directions include quantifying the social costs of infrastructure disruptions and better integration of equity into resilience decision-making.

## Introduction
The increasing scale, intensity, and frequency of disasters have revealed the fragility of infrastructure systems (World Meteorological 2021). In the last decade, disasters such as the Haiti Earthquake (2010), Tohoku earthquake and tsunami (2011), Hurricane Harvey (2017), the series of bushfires in Australia (2019) and wildfires in California (2017-2023), COVID-19 pandemic (2019-2023), and Winter Storm Uri (2021), along with ongoing drought in East Africa, to name a few, have significantly impacted communities. Beyond direct damage to homes and livelihoods, disasters also disrupt essential services provided by infrastructure systems, such as access to clean water, transportation, healthcare, or electricity. Critical infrastructure systems, or 'lifeline systems', play a vital role in maintaining and restoring the well-being of communities after hazardous events (Scherzer, Lujala et al. 2019). Unfortunately, current methods of infrastructure management do not adequately integrate the equitable recovery of communities, leading to erosion of trust and hampering long-term resilience.

Equity, in a broad sense, refers to the impartial distribution and just accessibility of resources, opportunities and outcomes which strive for fairness regardless of location and social group (Hart, 1974; Cook & Hegtvedt, 1983). In the realm of infrastructure management, equity entails addressing and mitigating systemic disparities to ensure universal distribution and access to the services provided by infrastructure. Recent studies have increasingly examined evidence of inequities in infrastructure service disruptions (Coleman, Esmalian et al. 2020). Literature has shown that socially vulnerable populations (such as those with low-income, minority, elderly, children, and people with health conditions or impairments) are disproportionately impacted by disasters (Flanagan, Gregory et al. 2011, Fatemi, Ardalan et al. 2017). They often face greater exposure, greater hardship, have less preparedness, and receive less institutional support to manage disasters (Fransen et al., 2023; Donner & Rodríguez, 2008 Fothergill, Maestas et al. 1999, Fothergill and Peek 2004, Rodríguez, Quarantelli et al. 2007). Along with other factors like settlement location and individual social vulnerability, disrupted infrastructure amplifies and extends these disparities (Dargin & Mostafavi, 2020, Dong et al., 2020). Research has shown an association between vulnerable groups facing more intense losses and longer restoration periods of infrastructure disruptions due to planning biases, inadequate maintenance, and governance structures (Hendricks and Van Zandt 2021). In addition, the network and interconnected structure of infrastructure systems can leave certain groups of people and certain areas more vulnerable to disrupted infrastructure (Esmalian, Coleman et al. 2021, Patrascu and Mostafavi 2023). Addressing inequities is a precursor to more resilient communities (Schlör, Venghaus et al. 2018), and research in the context of flood risk has argued to focus on welfare and well-being instead of asset losses, thereby putting distributive justice to the forefront of resilience (de Bruijn et al., 2022; Hallegatte & Walsh, 2021; van Hattum et al., 2020). A failure to understand equity can exacerbate disasters and lock inequality, which in turn, reduces resilience and leads to a vicious cycle (Boakye, Guidotti et al. 2022, Pandey, Brelsford et al. 2022).

However, the prevailing practices of infrastructure resilience have major limitations in the consideration of equity. Engineers often focus on prevention instead of resilient and equitable recovery (Coleman, Esmalian et al. 2020). First, infrastructure owners, managers, and operators tend to neglect the distinct social and cultural values of the community and their unique relationship to infrastructure systems (Hendricks and Van Zandt 2021). While infrastructure services are universally needed, the impact of infrastructure losses is not uniformly experienced by everyone. Second, the conventional approach to restoring infrastructure during hazard events is either based on the number of outages or the number of affected customers within an area, depending on the company preferences, not on priority needs or vulnerability. Although a seemingly intuitive and fair procedure, this approach overlooks the range of systematic disparities evident in infrastructure management and operation both in normal times and in hazard events. By focusing on the number of outages or affected customers, smaller or rural communities may be overlooked. Rural areas may have fewer alternatives and less redundancy if their system fails such as the disruption of a critical highway system. Vulnerable groups may have greater dependence on the infrastructure system such as elderly citizens who rely on powered medical devices and lower-income households that cannot afford power generators as substitutes. Finally, there are limited rules and guidelines that require infrastructure systems to consider vulnerable populations in restoration prioritization and resource allocation. Even as the general idea of equity is gaining traction, for instance around access to critical resources (Logan & Guikema, 2020), there are no standard procedures and methodologies to quantify and apply equity to infrastructure resilience processes. As such, infrastructure managers, owners, and operators are unlikely to recognize inequalities in service provision (Liévanos and Horne 2017). We argue that attention should shift to a resilience perspective: if and when infrastructures are disrupted, what are the mechanisms in place to ensure that *all* community members are well-provided and protected?

The field of equitable practices in infrastructure resilience management has sparked increasing interest over the last decade. Yet, while it is widely recognized that infrastructures impact equity, there is little knowledge of the mechanisms via which the design and planning of infrastructure systems impact the distributive effects of disasters. Studies have proposed frameworks to analyze the relationship of equity in infrastructure management (Clark, Tabory et al. 2022, Toland, Wein et al. 2023), adapted quantitative and qualitative approaches (Zhai, Peng et al. 2020, Yuan, Fan et al. 2021), and created decision-making tools for equity in infrastructure resilience assessment and management (Logan and Guikema 2020, Seigerman, McKay et al. 2022). With a decade of increasing interest in integrating equity into infrastructure resilience, it is now imperative to systematically evaluate the state of knowledge to understand the current progress and gaps. Hence, the purpose of this systematic literature review is to synthesize the growing body of literature of equitable thinking, research, and practices in infrastructure management as well as identify trends in the literature and the remaining knowledge gaps to fully integrate equity into measurable metrics. This literature review focuses on the intersection of equity, infrastructure, and natural hazard events through a systematic review process of over 3000 articles. The paper will focus on three overarching questions:

1. What are the current theoretical, conceptual, and methodological approaches to considering the disproportionate impacts of disrupted infrastructure services?
2. What are the similarities and differences among the studies (e.g. types of infrastructure, disaster, data)?
3. What are the current gaps of knowledge and future challenges of studying equity in infrastructure resilience?

This in-depth decadal review will bring insights into what aspects are fully known, partially understood, or completely missing in the conversation involving equity, infrastructure, and disasters. The paper will start with definitions of equity. It continues with the development of an eight-dimensional assessment framework that guides the analysis of literature. Significant results dive into the interactions between the dimensions of the framework. The review concludes with a vision statement that highlights the path forward for equity in infrastructure resilience and provides recommendations for future research.

## Definitions of Equity

The equity perspective has been neglected in conventional infrastructure resilience literature and practice, though several institutions have acknowledged the importance of integrating equity to ensure the high resilience and swift recovery of communities. The National Academies have stressed the significance of an equitable and resilient system to counteract the inherent discrimination and biases of infrastructure planning (National Academies of Sciences and Medicine 2022). The Federal Emergency Management Agency (FEMA) has attempted to reduce systemic barriers in mitigation practices (FEMA 2022). Here, we bring attention to how equity is being introduced into infrastructure management and which sources of literature served as inspiration for the four equity dimensions of distributional-demographic (D), distributional-spatial (S), capacity (C) , and procedural.

It is first important to distinguish between equal and equitable treatment which is often confused by research scholars. As stated by Kim and Sutley (2021), equality creates equivalence at the beginning of a  process whereas equity seeks equivalence at the end. Often, the term is interpreted through other social-economic concepts such as social justice (Boakye, Guidotti et al. 2022), sustainability (Seigerman, McKay et al. 2022), vulnerability (Karakoc, Barker et al. 2020), welfare (Silva-Lopez, Bhattacharjee et al. 2022, Dhakal and Zhang 2023), and environmental justice (Sotolongo, Kuhl et al. 2021). Equitable infrastructure is frequently associated with pre-existing inequities such as demographic features  (Atallah, Djalali et al. 2018, Coleman, Esmalian et al. 2023),  spatial clusters (Balomenos, Hu et al. 2019, Wakhungu, Abdel-Mottaleb et al. 2021), and political processes (Millington 2018).

As a burgeoning field, it is not surprising that there are few formal definitions of applying equity to infrastructure.  We summarize several definitions to understand how scholars are beginning to categorize equity in infrastructure management. Equity research has emphasized the attention to specific *demographic groups (D)* and *spatial areas (S)*. For instance, equitable infrastructure addresses the systemic inequalities in communities to ensure everyone has access to the same opportunity and outcome of infrastructure services (Coleman, Esmalian et al. 2023). Similarly, Clark, Tabory et al. (2022), define equity as the distribution of burdens and benefits with the goal of reducing disparity for the most disadvantaged populations. Silva-Lopez, Bhattacharjee et al. (2022) classify horizontal equity as impacts being distributed to groups deemed equal in ability and need while vertical equity is distributed among groups differing in ability and need.

According to Clark et. al (2018), the capabilities approach in equity connects the supporting infrastructure to the hierarchy of needs which recognizes the specific *capacities (C)* of households. Inspired by the environmental justice literature, Seigerman, McKay et al. (2022) also details recognitional equity (acknowledging and respecting human social differences); procedural equity (inclusive participation and transparent planning); and distributional equity (access to resources or exposure to environmental harms). Rendon, Osman et al. (2021) calls back to recognitional and *procedural equity (P)* by stating that the validity of local cultural identities is often overlooked in the participation process of designing infrastructure.

Based on the existing literature and terminology, the systematic literature review proposes four dimensions of equity: distributional-demographic (D), distributional-spatial (S), procedural (P), and capacity (C) equity. We acknowledge recognitional equity as the root system to anchor diverse thinking in equity; however, these principles are not prominently represented in the collected literature, and therefore it is not included in our review framework.

**Distributional-Demographic (D)** equity represents accessibility to and functionality of infrastructure services considering the vulnerability of demographic groups (Beck and Cha 2022). Achieving distributional-demographic equity means reducing or eliminating disparate access to goods, services, and amenities among different populations such as racial and ethnic minorities, lower socioeconomic status, children and elderly, those who are disabled and with chronic health conditions, among others (Dhakal, Zhang et al. 2021). These groups may need greater support due to greater hardship to infrastructure losses, greater dependency on essential services, and disproportionate losses to infrastructure (Coleman, Esmalian et al. 2020, Esmalian, Wang et al. 2021, Toland, Wein et al. 2023).

**Distributional-Spatial (S)** equity focuses on the equitable distribution of infrastructure services to all spatial regions (Matin, Forrester et al. 2018). Inequitable access may be due to the optimization and operation of the system which may leave certain areas in isolation (Adu-Gyamfi, Shaw et al. 2021, Fan, Jiang et al. 2022, Best, Kerr et al. 2023). An equitable access to essential services (EAE) approach to spatial planning can identify service deserts (Logan and Guikema 2020). Urban and rural dynamics may also influence infrastructure inequities. Often, rural areas have deficient funding sources compared to urban areas (Pandey, Brelsford et al. 2022).

**Procedural (P)** equity is the inclusion of all individuals in the decision-making process from the collection of data to the influencing of policies. According to Rivera (2022), inequities in the disaster recovery and reconstruction process originate from procedural vulnerabilities associated with historical and ongoing power relations. Governments and institutions may have excluded certain groups from the conversation to understand, plan, manage, and diminish risk in infrastructure (Eghdami, Scheld et al. 2023). As argued by Liévanos and Horne (2017), such utilitarian bureaucratic decision rules can limit the recognition of unequal services and the development of corrective actions. These biases can be present in governmental policies, maintenance orders, building codes, and distribution of funding sources (Kim and Sutley 2021).

**Capacities (C)** equity is the ability of individuals to counteract or mitigate the effect of infrastructure loss. As mentioned by Parsons, Glavac et al. (2016), equity can be enhanced through a network of adaptive capacities at the household or community level. In resilience literature, these adaptive capacities are often viewed as the backbone of community resilience (Champlin, Sirenko et al., 2023). In the case of infrastructure, households can own service substitutes such as power generators or water storage tanks (Stock, Davidson et al. 2021, Abbou, Davidson et al. 2022). Based on household decision-making behaviors, capacity can be applied in different stages of the disaster including preparation, mitigation, and response. (Esmalian, Wang et al. 2021). However, capacity can be limited by people's social connections and social standing.

## Methods of Systematic Literature Review

*Document Search and Filter:* Our systematic literature review followed the PRISMA protocol, as highlighted in Figure 1. The search covered Web of Science, Science Direct, and Google Scholar. To cover a broad set of possible disasters and infrastructures, our search focused on the key areas of equity ("equit- OR fair- OR justice- OR and access-"), infrastructure ("AND infrastructure system- OR service-"), and disasters (" AND hazard- OR, cris- OR, disaster- OR"). We limited our search to English publications from the engineering and social sciences during January 2010 to March 2023. Using the COVIDENCE program, we had a minimum of 2 people approve each article. We excluded literature reviews and opinion pieces. Excluding duplicates, 2,990 articles were screened based on their title and abstract which yielded 397 full-text studies. From here, 277 studies were excluded based on insufficient equity discussion, limited infrastructure focus, limited disaster focus, wrong study design, and unable to access which yielded 120 final articles.

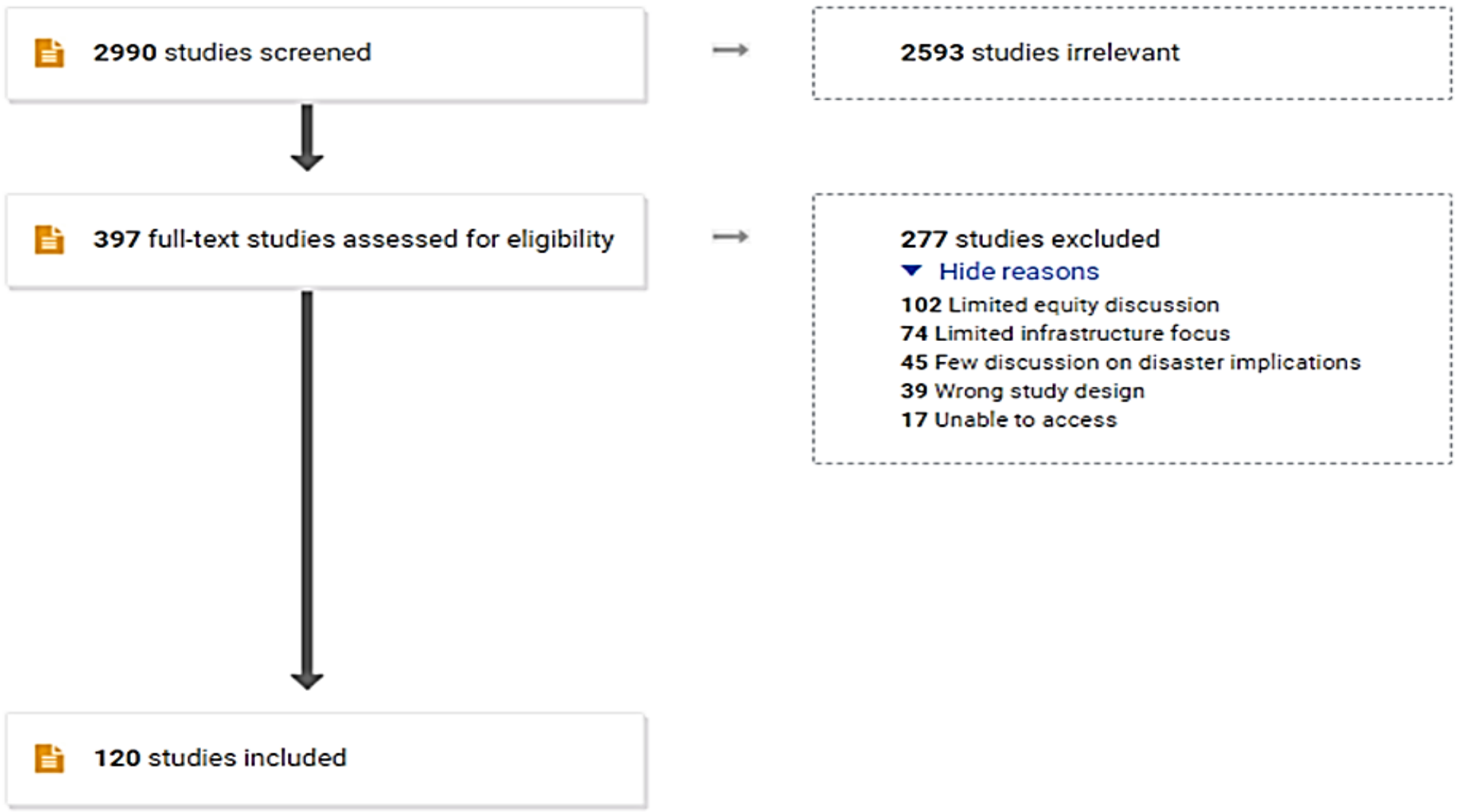

**Figure 1.** Filtering of systematic literature review, generated using PRISMA on COVIDENCE

*Assessment Framework:* We designed an eight-dimensional assessment framework (see Figure 2) to analyze the literature and answer our three overarching research questions. The framework distinguished the *focus of the study* (infrastructure type, geographical scale, geographic location, temporal scale, hazard event type); the *methodological approaches* (nature of study/ data collection and theoretical perspective), and *equity dimensions* addressed (Figure 2). In the following, we elaborate on these concepts and sub-questions as well as the inclusion and exclusion criteria of the dimensions.

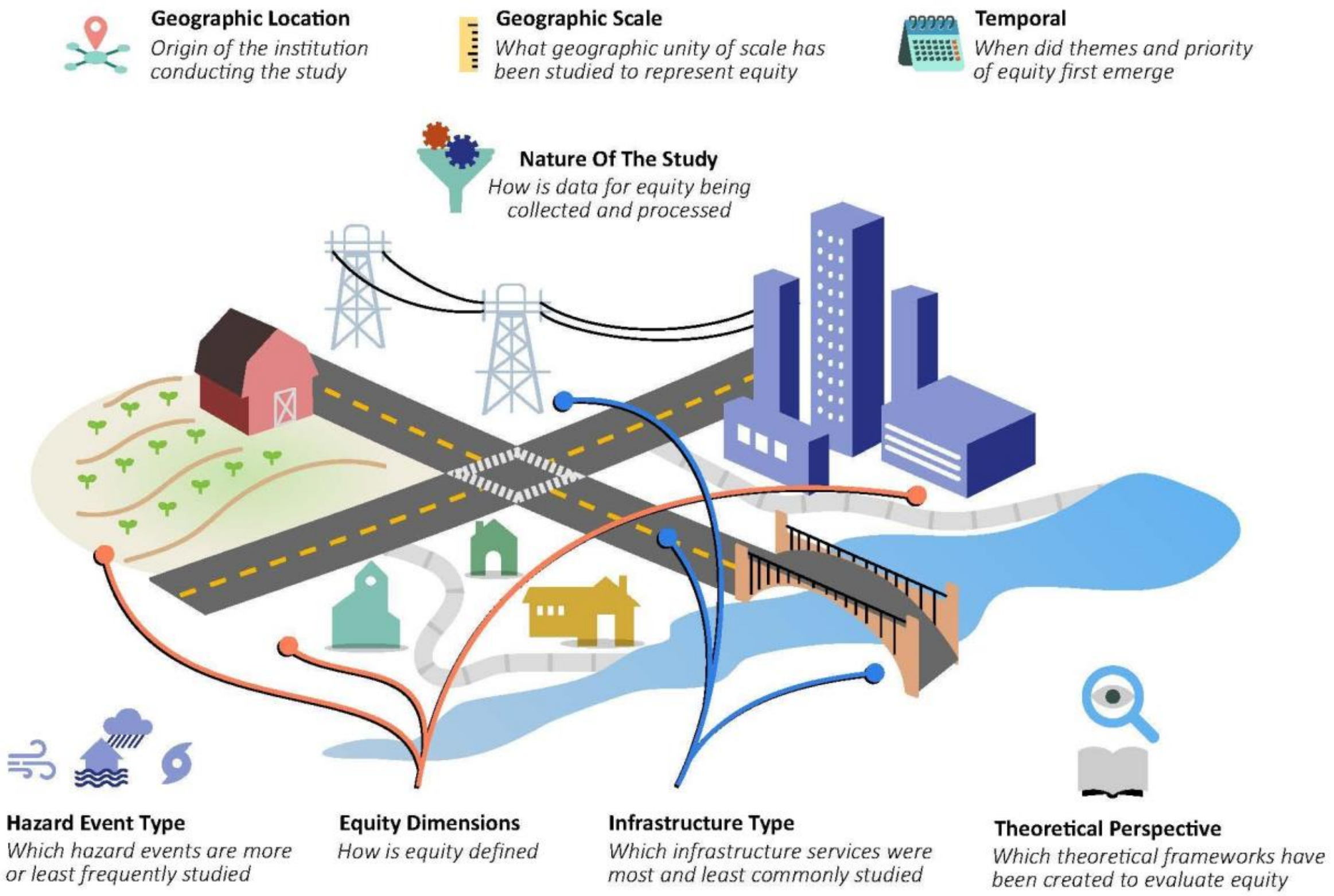

**Figure 2.** Eight-dimensional assessment framework to evaluate studies related to equity in infrastructure resilience

Infrastructure Type: *Which infrastructure services were most and least commonly studied?* This category is divided into power, water, transportation, communication, health, food, sanitation, stormwater, emergency, and general if a specific infrastructure is not mentioned. It also lists whether studies investigated one or multiple systems. Green infrastructure, building structures, and industrial structures were excluded given that this literature review focuses on critical infrastructure.

Nature of the Study: *How is data for equity being collected and processed?* This category analyzed data types used including conceptual, descriptive, open-data, location-intelligence, and simulation data. To clarify, conceptual refers to purely conceptual frameworks or hypothetical datasets; descriptive refers to surveys, questionnaires, interviews, or field observations; open-data refers to any open-data source that is easily and freely attainable such as census and flood data; location-intelligence refers to social media, human mobility, satellite and aerial images, and visit data and typically requires large computational and monetary expenses; and finally, simulation data can be developed through simulation models or Monte-Carlo simulation. Second, the data can be processed through quantitative or qualitative methods. Quantitative methods may include correlation, principal component analysis, and spatial regression while qualitative methods may include thematic coding, participatory rural appraisal, and citizen science. When studies had several analyses, the most common and significant ones were included. For example, it can be assumed that studies of linear regression discussed correlation analysis and other descriptive statistics in their data processing.

Geographic Location: *Which countries have studied equity the most and least?* This category is at the country scale such as the United States and other countries like the Netherlands, China, and Australia, among others.

Geographic Scale: *What geographic unit of scale has been studied to represent equity?* Smaller scales of study can reveal greater insights at the household level while larger scales of study can reveal comparative differences between regional communities. It ranges from individual, local, regional, system, and country. To clarify, individual can include survey respondents, households, and stakeholder experts; local is census block groups, census tracts, and ZIP codes equivalent scales; regional is counties, municipalities, and cities equivalent scales; and system focuses on the nodes and network outline.

Temporal scale: *When did themes and priority of equity first emerge?* First, this category determines when equity in infrastructure research is published and whether these trends are increasing, decreasing, or constant. Second, it evaluates at what stage equity is applied in infrastructure management with respect to the disaster management cycle. For example, equity could be applied before a disaster, in the immediate aftermath, or in long-term recovery monitoring.

Hazard Event Type: *Which hazard events are most or least frequently studied?* This category includes flood, tropical cyclone, drought, earthquake, extreme temperature, pandemic, and general if there is no specific hazard. To clarify, tropical cyclones include hurricanes, typhoons, and tsunamis while extreme temperatures are winter storms and heatwaves. It determines which studies are specific to hazards and which can be applied to universal events.

Theoretical Perspective: *Which theoretical frameworks have been created and used to evaluate equity?* This category summarizes the reasoning behind the theoretical frameworks which may have informal or formal names such as a service-gap model, well-being approach, and capability approach.

Equity Dimensions: *How is equity conceptualized and measured?* First, the category evaluates whether the studies are directly studying equity in infrastructure or if equity and infrastructure are pieces of an overall resilience analysis. Second, it summarizes the equity reasoning and significant equity conclusions. Third, it labels the equity into four dimensions (DPSC).

**Results**

The results reveal the spatiotemporal patterns of equity, data collection of equity, methods to analyze equity, the most common pathways between infrastructure, hazard, and equity, as well as the most prominent gaps. Figures 1A-12A provide additional context to the research findings and can be found in the Supplemental Information.

*Spatiotemporal patterns of equity:* Overall, there is an increasing number of publications about equity in infrastructure management (Figure 3). There is a rising trend towards directly examining equity considerations in disrupted infrastructure (in blue) with varying levels of an overall resilience framework with infrastructure and equity components. A slight decrease observed in 2021 could be because of the focus on COVID-19 research.

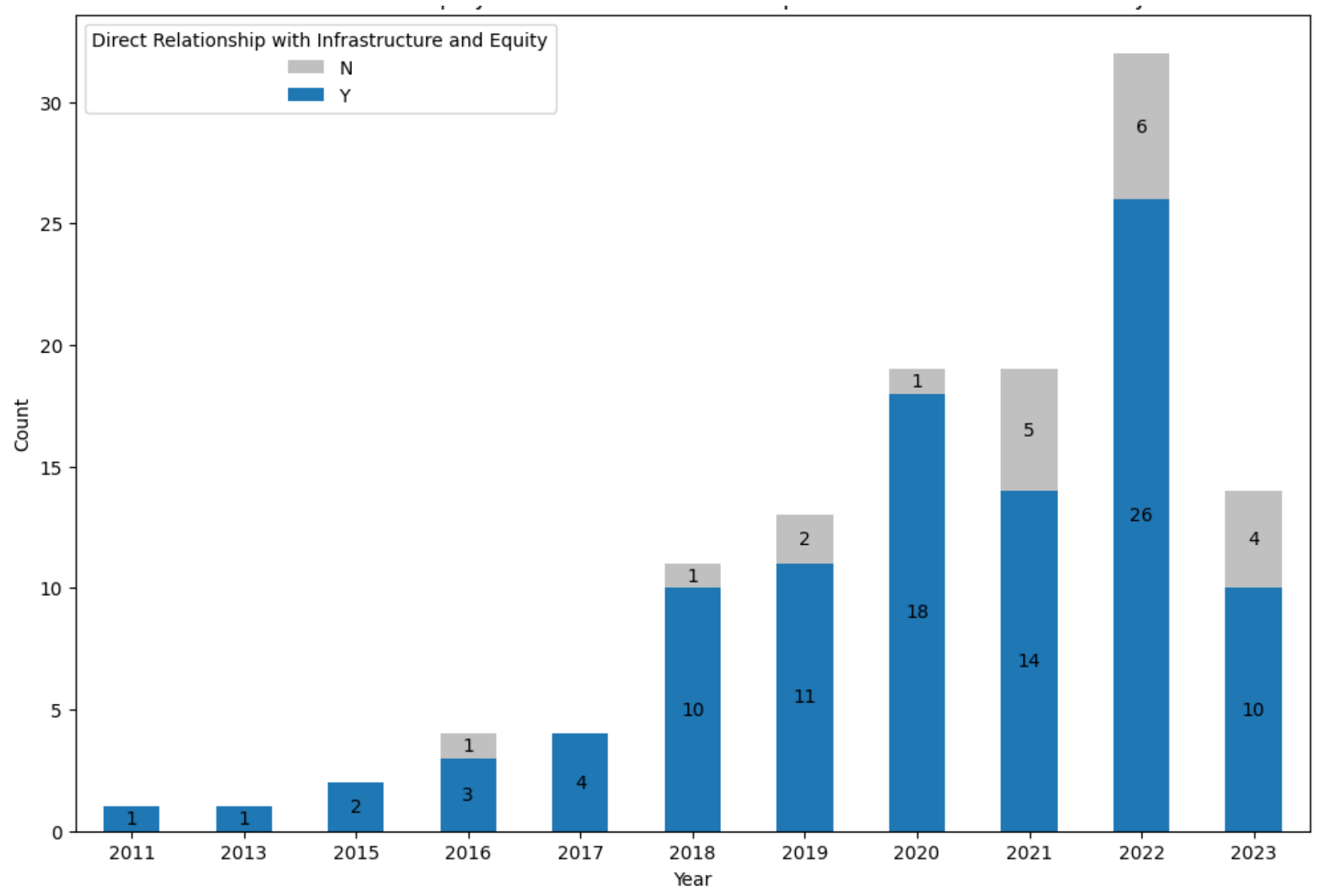

**Figure 3.** Publishing frequency of equity in infrastructure management journal articles. The stacked bar chart shows the frequency of studies that included equity and infrastructure as pieces for overall resilience (grey) and those which directly studied equity (blue). Note that only January–March is included in 2023.

Spatially, most studies focus on the US (61.67%) (Figure 4) along with the global north (70%) (Figure 11A). Figure 4 also highlights the growing trend towards more integrated frameworks.

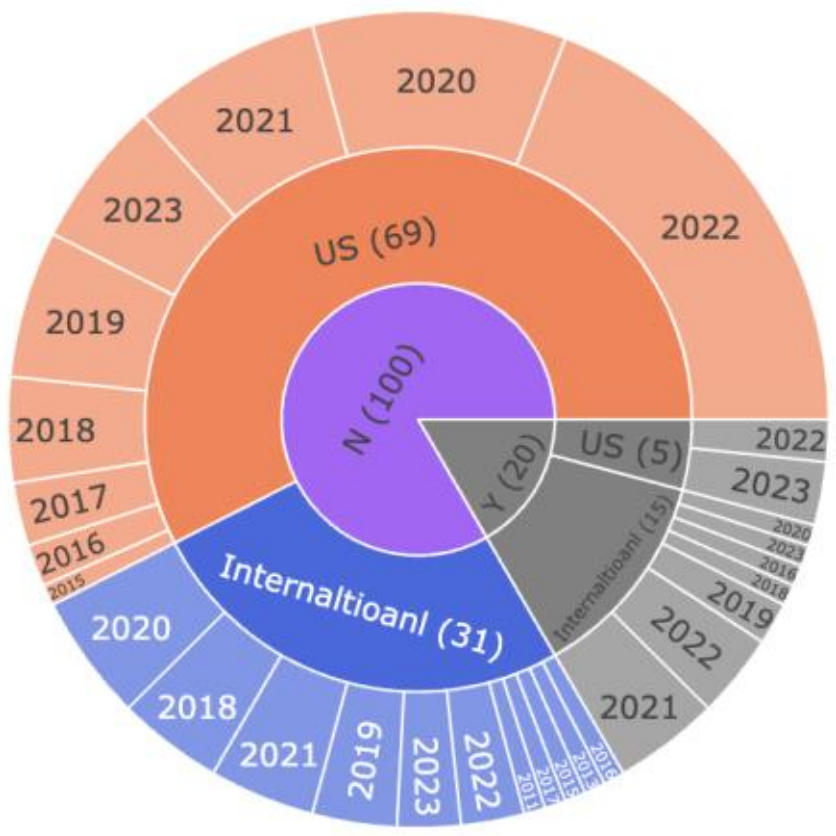

**Figure 4.** Distribution of US and International studies given the integration of equity and over the years

Equity can be applied in different points of infrastructure management. Based on the general trends in the literature, we identified specific timescales and connected them with the standard disaster management cycle. Mitigation includes recognition of ideas from various stakeholders and investment and upgrades in infrastructure. Preparedness includes the development of decision-making plans and resource stockpiling of essential services. Response includes emergency response plans enacted by utility companies and emergency managers, and resource deployment is for the temporary restoration of services. Recovery includes both short-term infrastructure restoration and long-term sustainability to prevent a recurrence of a disruption.

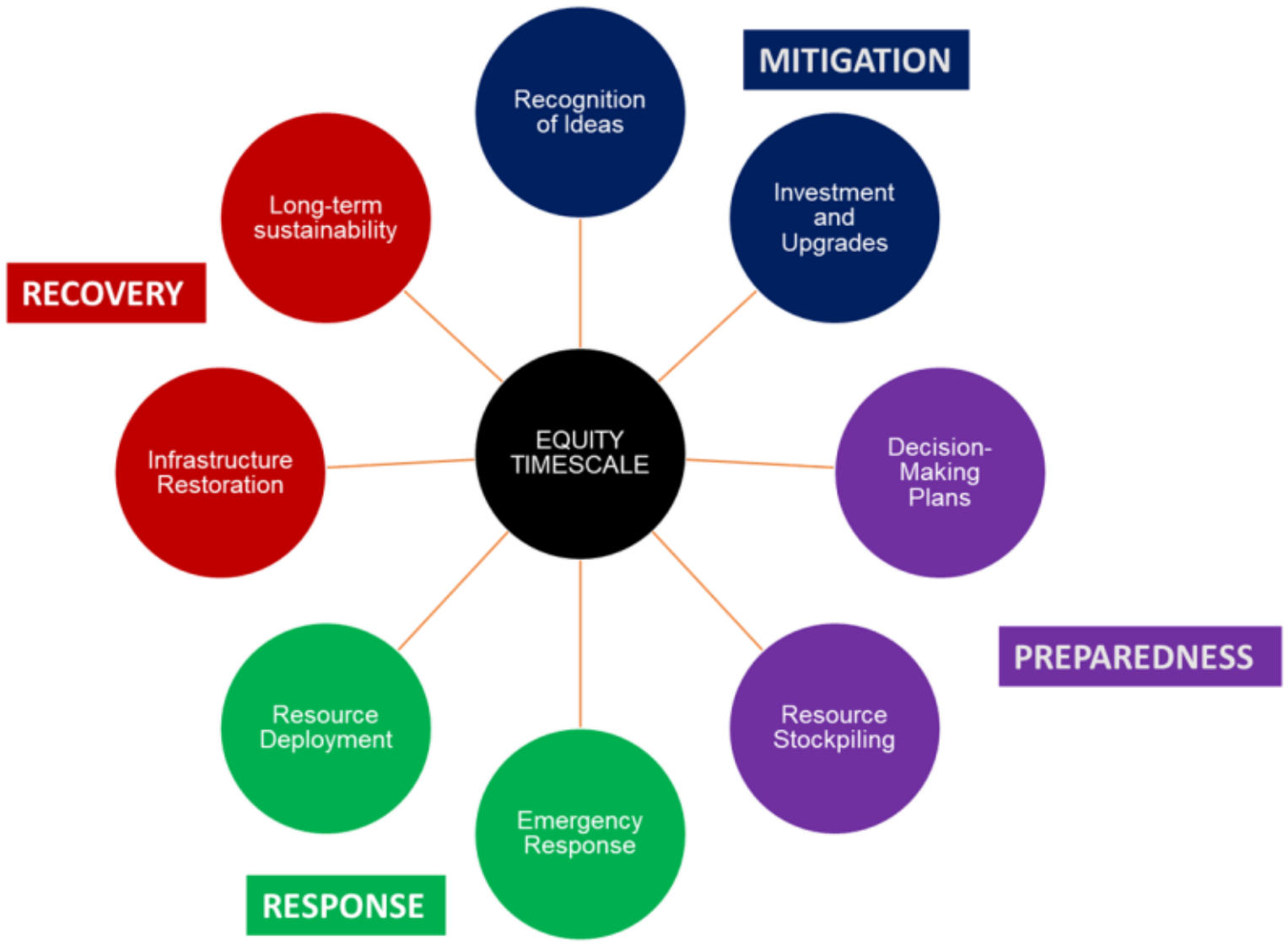

**Figure 5.** Applying equity at different timescales in relation to the disaster management cycle

*Data Collection of Equity:* Our Sankey diagram (Figure 6) sketches the distribution of data collection pathways. Most studies start from quantitative data (141), with only few using qualitative (14) or mixed data (17). Most quantitative studies use open data (60), whereas descriptive data (58) integrates quantitative, qualitative, and mixed data collection. The most prominent scale is the individual/household scale (61), largely stemming from descriptive data collected from surveys (Figure 6). In contrast, open data fed largely into the system, local, and regional level spatial scales. Fewer studies used location-intelligence (24) and simulation data (26). We also noted that international studies do not use any location-intelligence data which could be due to limitations in data availability and different security restrictions to these researchers.

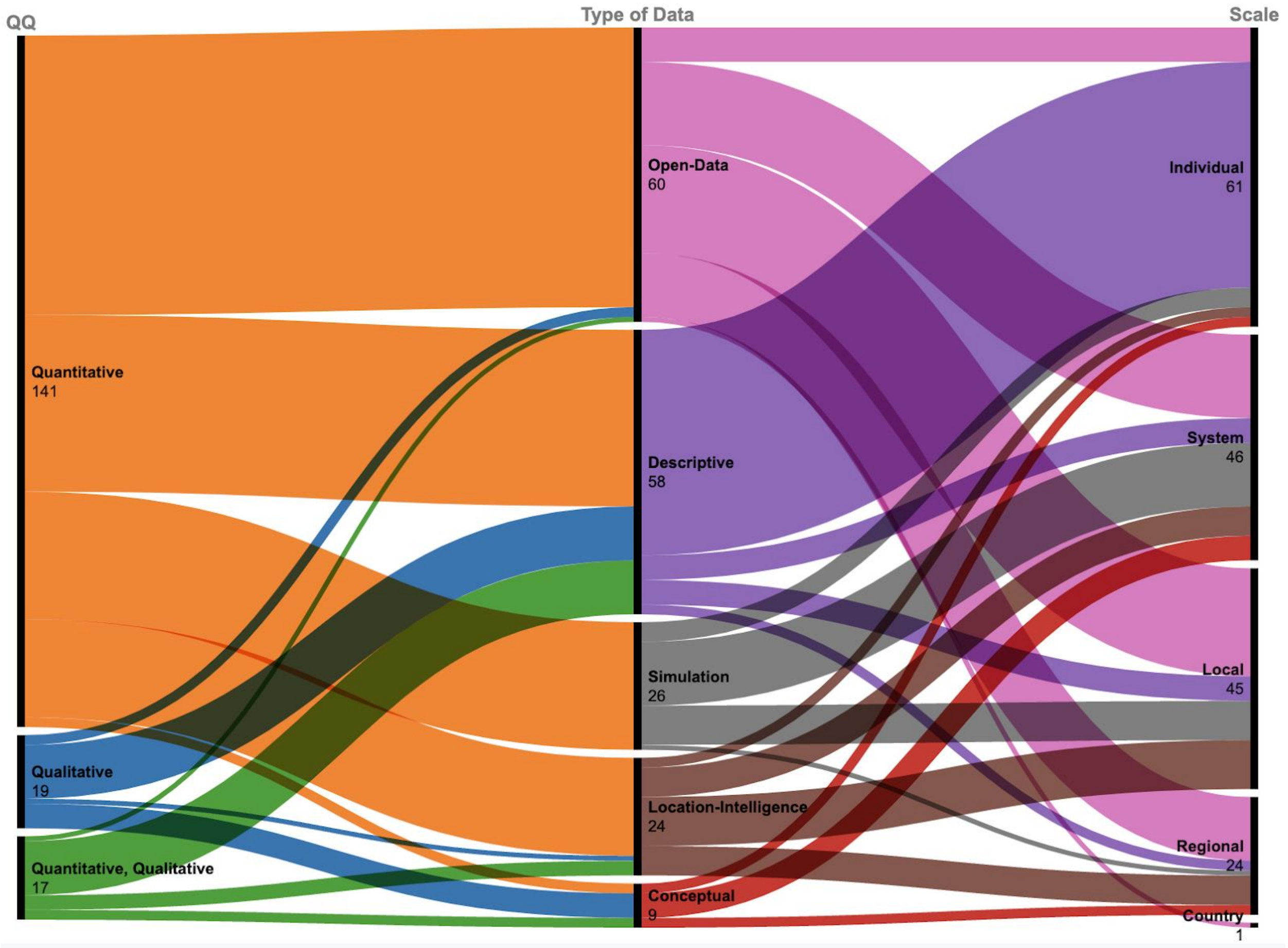

**Figure 6.** Sankey diagram that flows the quantitative-qualitative, data type, and scale of analysis

Figure 7 highlights how the different data sources relate to the conceptualization of equity. The most common approach is using descriptive data for distributional-demographic (D) equity (Figure 7). For each data type, distributional-demographic (D), in fact, is the dominant equity dimension. Distributional-spatial (S) is equal for open-data (13) and location-intelligence (13) which connects to the multitude of spatial data availability at coarser and finer scales, respectively. Comparatively, procedural (P) and capacity © equities are limited in the current literature (<10); however, there is a moderate pathway for capacity equity using descriptive-quantitative data (19).

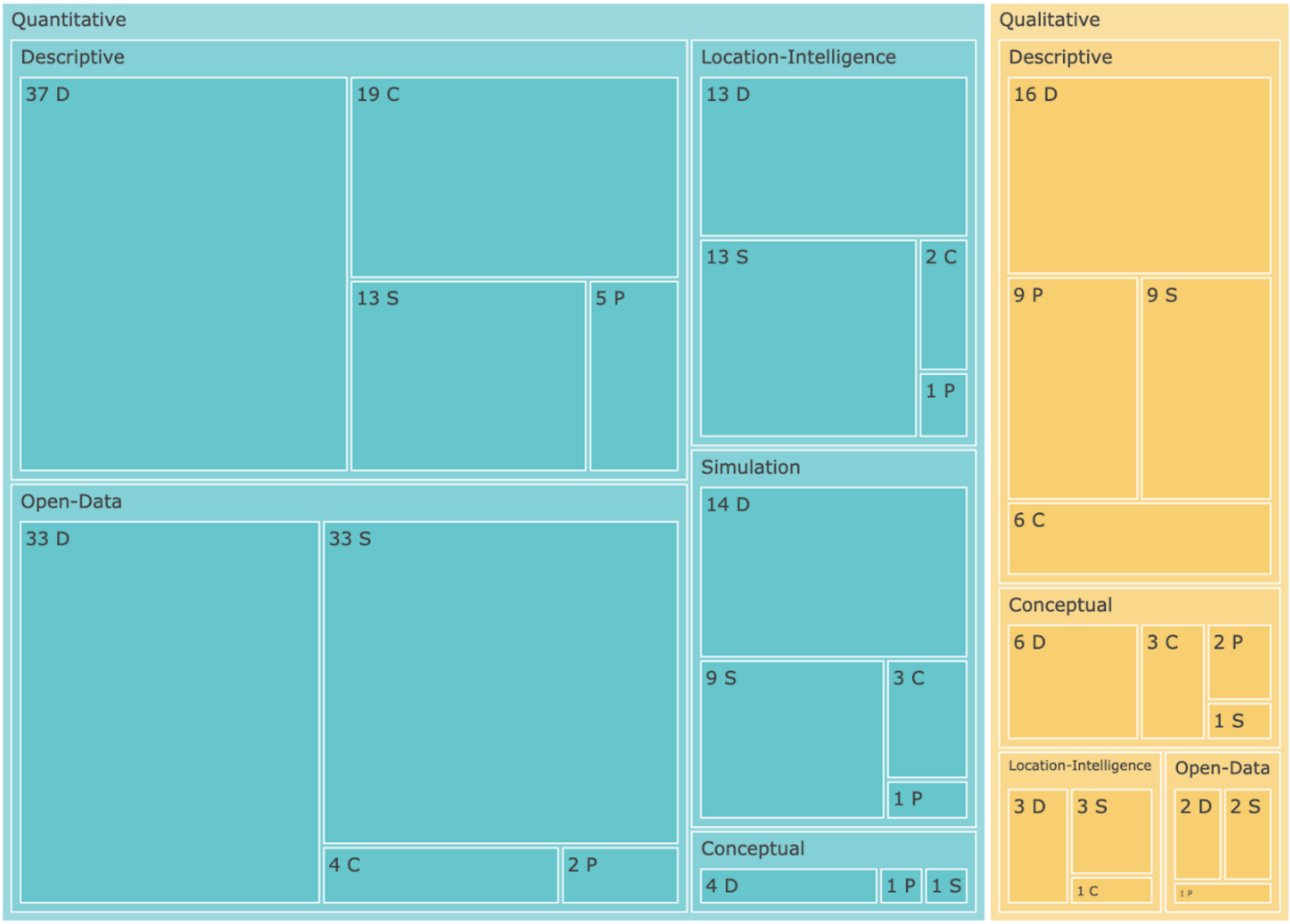

**Figure 7.** Tree diagram of the quantitative-qualitative, data type, and equity dimension

A major problem with quantitative data collection for resilience is digitally invisible populations (Longo et.al 2017; Sung, 2015). Examples of digitally invisible populations may include groups that are not properly represented in digital media. Location-intelligence data, especially, relies on cell phone pinpoint data, and certain vulnerable groups may not afford or have frequent access to cell phones. To avoid overlooking especially vulnerable communities that may not leave a digital footprint, we discuss the concept of equitable data. As stated by Gharaibeh, Oti et al. (2021), equitable data represents *all* communities in the study area. Very few studies call into question the fairness of the data collection to adequately represent the multi-dimensions of equity (Yuan, Fan et al. 2021). Regarding location-intelligence data, there is also a trade-off between proper representation of demographic groups and ensuring the privacy of individuals (Dhakal, Zhang et al. 2021, Yuan, Esmalian et al. 2022). Location-intelligence data has captured the accessibility to essential services (Yuan, Esmalian et al. 2022) and public emotions to disruptions (Chen and Ji 2021, Batouli and Joshi 2022). Further, satellite information (Roman, Stokes et al. 2019), telemetry-based data (Coleman, Esmalian et al. 2023), and human mobility data (Lee, Maron et al. 2022) were used to evaluate the equitable restoration of power systems and access to critical facilities.

*Methods to Interpret Equity*: As shown in the Supplemental Information of Figure 6A and Figure 7A, there are distinct quantitative and qualitative methods to interpret equity. Regarding quantitative analysis, the top methods were the descriptive statistics of correlation (13), principal component analysis (PCA) (9), chi-square (7), and ANOVA (6) mean along with spatial analysis of geographic information system (GIS) (16) and Moran's-I statistic (9). It also included statistical models of Monte Carlo (9), logit (13), and ordinary least squares (OLS) (5). Most quantitative methods were focused on descriptive analysis and linear models which can assume simple relationships within equity dimensions. Thus, more complex models are needed to uncover the underlying mechanisms associated with equity in infrastructure.

Complex models such as agent-based, decision trees, and simulation, could unearth non-linear relationships between variables and equity in infrastructure. For instance, Esmalian, Wang et al. (2021) used agent-based modeling to incorporate the social demographic features and their mitigation responses to power outages caused in Hurricane Harvey. In another agent-based model, Baeza, Bojorquez-Tapia et al. (2019) considered the tradeoff of three policies for infrastructure investment. The policies were prioritizing neighborhoods of high social pressure, building new access in places without infrastructure, and repairing aged infrastructure. Using a classification and regression tree model, Dargin and Mostafavi (2022) created hardship pathways to infrastructure outages based on sociodemographic and preparedness features. Still, other studies have leveraged simulation models to understand access to water (Toland, Wein et al. 2023), health (Dong, Esmalian et al. 2020), and transportation (Silva-Lopez, Bhattacharjee et al. 2022). Additionally, conceptual studies are proposing methods such as gravity-weighted models (Clark, Peterson et al. 2023) and genetic algorithms (Kim and Sutley 2021).

Regarding qualitative analysis, the most common methods were thematic coding (6), participatory rural appraisal (2), citizen science (2), sentiment (2), content and document analysis (2), intra-urban comparative risk (1), and photovoice (1). The Supplemental Information shows the variety of methods used across the four dimensions (Figure 6A). Qualitative methods are essential to capture diverse angles of equity such as unexpected strategies and coping mechanisms used in capacity equity (Daramola, Oni et al. 2016, Kohlitz, Chong et al. 2020) or the perspectives of stakeholders used in procedural equity (Islam, Shetu et al. 2022, Masterson, Katare et al. 2023). It can also narratively explain the personal hardships experienced by people (Stough, Sharp et al. 2016, Gbedemah, Eshun et al. 2022).

*Representation of Equity Dimensions:* Figure 8 shows how the different dimensions of equity relate to the different infrastructure systems under investigation. In total, a combination of 715 pathways with overlapping studies were included. Referencing the infrastructure systems, studies with power (165), water (144), and transportation (107) were the most frequent while studies with stormwater (20) and emergency (9) services were the least frequent. Referencing demographics, the most studied were income (146), ethnicity (118), and age (118) while least studied were gender (63), employment (36), and intergenerational (1), which is in line with the findings around the dominance of distributional-demographic equity concepts that we reported before (Figure 7).

The finding of the papers, however, confirmed that *all* variables are generally connected with greater inequity; for instance, lower-income and minority households faced greater exposure, more hardship, and less tolerance to withstand power, water, transportation, and communication outages during Hurricane Harvey (Coleman, Esmalian et al. 2020, Coleman, Esmalian et al. 2020). In another study, Stough, Sharp et al. (2016) identified a lack of mobility, or transportation, as a frequent issue for respondents with disabilities after Hurricane Katrina. Islam, Shetu et al. (2022) focused on infrastructure losses that catered to women's needs such as family planning and maternal health issues. Others concluded that women were burdened more by infrastructure losses as they were expected to "pick up the pieces" and replace the disrupted services (Dominelli 2013, Sam, Abbas et al. 2021). A common point of conversation was that demographic vulnerabilities are interconnected and compounding, and often, distributional-demographic equity is a pre-existing inequality condition.

Despite the variety of social variables, most publications studied the socioeconomic status and minority households; few studies examined the involvement of indigenous populations (Ahmed, Kelman et al. 2019). Also, those that examined households with young children typically did not account for the direct impacts on children and instead concentrated on the impacts of household caretakers (Chakalian, Kurtz et al. 2019). Only one study discussed the concept of intergenerational equity. According to Lee and Ellingwood (2015), “intergenerational discounting methods can achieve equitable risk sharing between current and future generations.” The study found that even a slight difference in discounting rate can lead to vastly different decisions. Insufficient investments in design and planning will only increase the cost and burden of maintenance and replacement for future generations.

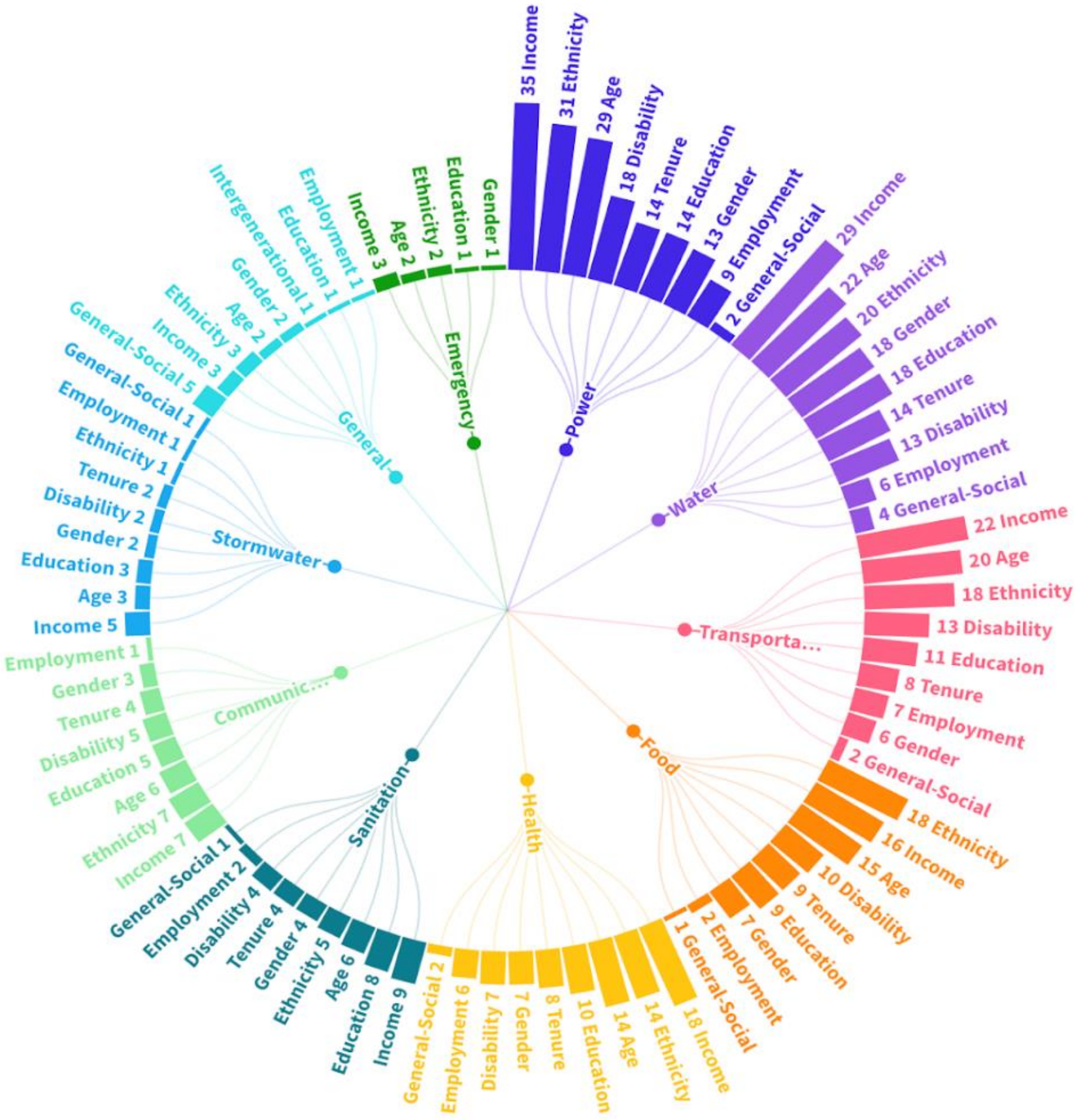

**Figure 8.** Flow diagram to reveal the connection between disruptions in infrastructure services and distributional-demographic equity

*Spatial Aspect of Equity:* Distributional-spatial equity was the second most studied dimension which includes spatial grouping and urban-rural designation, particularly given the rise of open-data and location-intelligence data with spatial information. Figure 9 shows how spatial aspects are represented across different infrastructures. In total, 107 pathways were found with spatial (86) and urban (21) characteristics. Power (25), transportation (22), and health (16) systems were the most studied with stormwater (5), emergency (3), and communication (3) the least studied. Urban-rural studies on communication and emergency services are entirely missing.

Several studies used Moran's I statistic to identify areas of high physical and social vulnerability (Ulak, Kocatepe et al. 2018, Dong, Esmalian et al. 2020, Esmalian, Coleman et al. 2021). Esmalian, Coleman et al. (2022) defined equitable access to grocery stores using spatial network attributes such as number of unique stores visited, average trip time, and average distance. Logan and Guikema (2020) identified the terms access rich and access poor, finding that White populations had less distance to travel to open supermarkets and service stations. As studied by Patrascu and Mostafavi (2023), vulnerable communities can be indirectly impacted by spatial spillover effects from neighboring areas. Regarding urban- rural studies, Pandey, Brelsford et al. (2022) argue that inequalities emerge when urban infrastructure growth lags with respect to urban population while rural areas face infrastructure deficits. Rural municipalities are often found with fewer resources, longer restoration times, and less institutional support (Mitsova, Esnard et al. 2018, Hamlet, Kamui et al. 2020, Kohlitz, Chong et al. 2020).

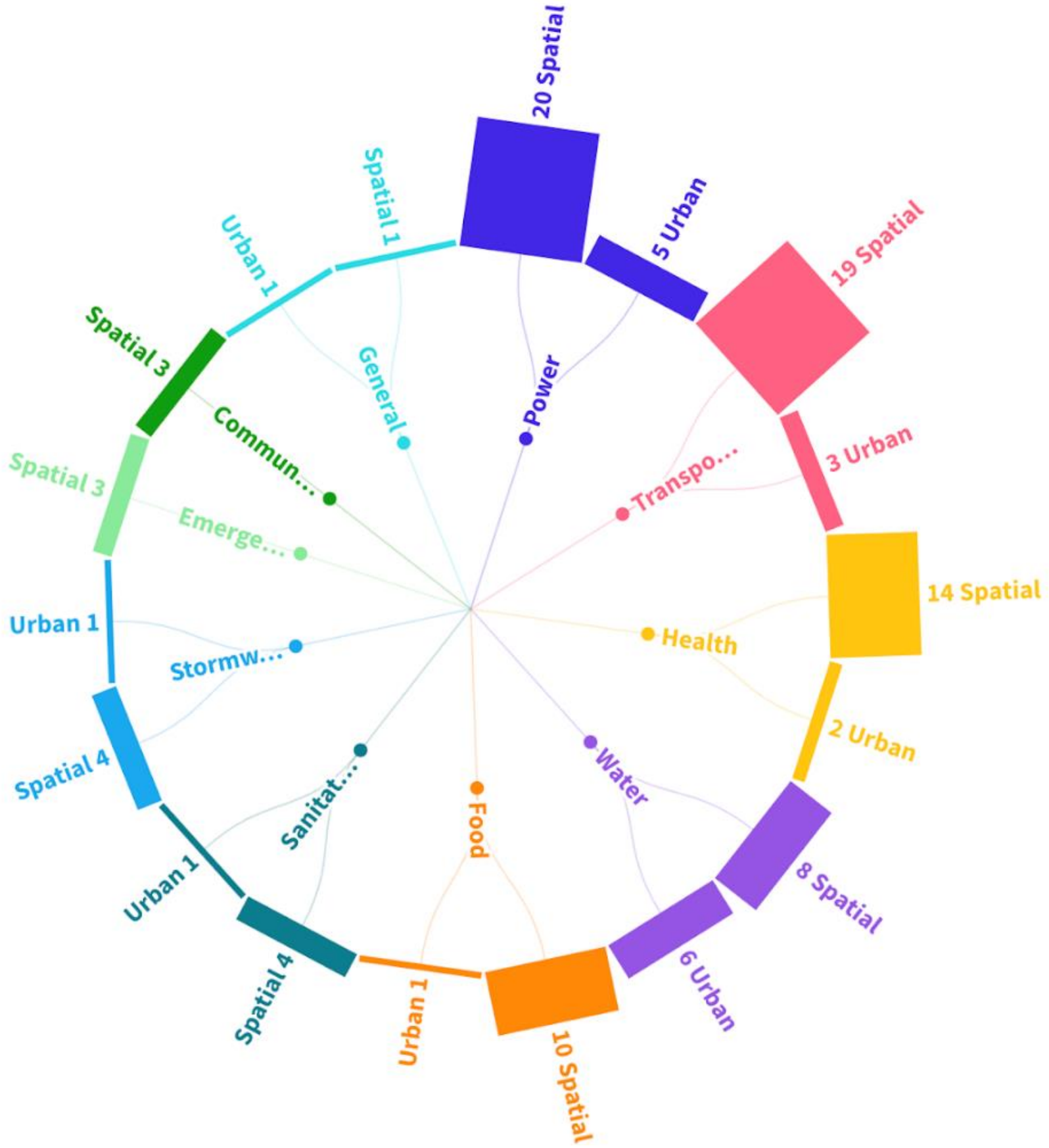

**Figure 9.** Flow diagram to reveal the connection between infrastructure services and distributional-spatial equity

*Building procedural and capacity equity:* Procedural and capacity equities were the least studied equity dimensions; however, they both play a critical role in community resilience. Regarding procedural equity, citizen science can incorporate environmental justice communities into the planning process, educate engineers and scientists, and collect reliable data (Hendricks et. al, 2018, Oti, Gharaibeh et al. 2019, Gharaibeh, Oti et al. 2021). It is also important to interact with invested stakeholders and coordinate across sectors (Yu and Welch 2022). Masterson, Katare et al. (2023) worked with Rockport, TX planners to interweave equity in their policy planning such as providing accessible services and upgrading infrastructure for lower-income and racial-ethnic minority neighborhoods. Additionally, Chen and Ji (2021) used sentiment analysis to capture public demand urgency which can assist with resource allocation and speed of restoration. To counteract disparities, another solution is to share local knowledge and better facilitate conversations between government and public (Johnson, Edwards et al. 2018). In fact, Baeza, Bojorquez-Tapia et al. (2019) acknowledged that choices regarding infrastructure investment can be influenced by less technical criteria such as electoral constituencies based on voting patterns or the need to respond to specific events with high political visibility.

Capacity equity can refer to service substitutes and coping strategies. For example, Chakalian, Kurtz et al. (2019) found that white respondents were 2.5 more likely to own a power generator than others while Kohlitz, Chong et al. (2020) discovered that poorer households could not afford rainwater harvesting systems. Coping strategies may include tolerating the disruption (Esmalian, Dong et al. 2021), cutting back on current resources (Daramola, Oni et al. 2016), or performing household adaptations (Abbou, Davidson et al. 2022) until the previous level of service is restored. A capabilities approach can also refer to the adjustment in time and money to achieve the same level of resources (Clark, Peterson et al. 2023). As noted by Rendon, Osman et al. (2021), communities that are impacted by racism and classism injustices are less able to build capacity. As such, equity can be promoted through community engagement and education programs (Atallah, Djalali et al. 2018).

*Intersection of infrastructure, hazard, and equity:* Tropical cyclones (34.8%) and floods (29.6%) make up over half of the studied hazards (Figure 4A) while power (21%), water (19%), and transportation (15.2%) were the most frequently studied infrastructure services (Figure 5A). The most common pathway across these infrastructure services was a tropical cyclone and flooding with distributional- demographic equity (Figures 7A-9A).

However, there are multiple approaches to addressing equity that we found in our review. For example, Ulak, Yazici et al. (2020) detected evidence of Black households having lower robustness compared to White households as well as lower income populations having a higher percentage of power losses during Hurricane Hermine. In testing the equity of alternative policies, the study found that the renewal of power network components was favored over increasing response crew sizes. Another example is the disproportionate experiences of water availability in a drought setting (Millington 2018). The study found that water scarcity was exacerbated by both the inequities of the water infrastructure and differentiated abilities of residents to store water in small-scale reservoirs. Several willingness-to-pay (WTP) models have been created to understand the relationship between households, infrastructure, and hazards. For example, higher WTP was associated with higher income in power, water, and transportation services (Wang, Sun et al. 2018, Stock, Davidson et al. 2022).

As an exploration of the frameworks, we selected the following and their applications to infrastructure resilience, hazard, and equity (Table 1). The full list of frameworks can be found in the excel database. Surprisingly, there is a very broad range of frameworks used. Some frameworks have their roots in capabilities approach (Clark, Seager et al. 2018, Abbou, Davidson et al. 2022), specifically the ability to access resources (Clark, Peterson et al. 2023, Dong, Esmalian et al. 2020). Additional studies used resilience or vulnerability assessment (Toland, Wein et al. 2023, Coleman, Esmalian et al. 2020) or are rooted in welfare economics (Dhakal and Zhang 2023, Ulak, Yazici et al. 2020). Others involved stakeholder and expert opinion to enhance community engagement (Yu and Welch 2022).

Several frameworks focused on household capability to infrastructure losses. For example, the service gap model introduced the zone of tolerance to encapsulate the different capabilities of households to power outages (Esmalian, Dong et al. 2021). Lower-income households were less likely to own power generators. According to Abbou, Davidson et al. (2022) households held specific adaptations to mitigate household losses. Women were more likely to use candles and flashlights while people with higher education were more likely to use power generators. Building on capability, other frameworks focused on the accessibility to resources. Clark, Peterson et al. (2023) developed the social burden concept which uses resources, conversion factors, capabilities, and functioning into a travel cost method to critical resources. In an integrated physical-social vulnerability, Dong, Esmalian et.al 2020 calculated disrupted access to hospitals.

Grounded in vulnerability assessments, Toland, Wein et al. (2023) considered an earthquake scenario to emergency food and water resources. The transportation justice threshold index framework used GIS to integrate social vulnerability into transportation understanding (Oswald Beiler and Mohammed 2016). Others described welfare economics to infrastructure losses. In the social welfare-based infrastructure resilience assessment, the study adapted the Gini coefficient in measuring the inequality of disaster impacts (Dhakal and Zhang 2023). Stock, Davidson et al. (2022) evaluated the willingness-to-pay approach and unhappiness measurements to understand inequities in different demographic groups.

Still, additional studies were based on stakeholder input and applicability. In a health-focused qualitative study, Atallah, Djalali et al. (2018) established an ABCD roadmap which included acute life-saving services, basic institutional aspects for low resource settings, community driven health initiatives, and disease specific interventions. Another example of including expert opinion was the development of the social resilience tool for water systems which used Delphi survey responses (Sweya, Wilkinson et al. 2021).

**Table 1.** Summary of theoretical frameworks as it relates to infrastructure, hazard, and equity

| **Citation** | **Framework** | **Infrastructure** | **Hazard** | **Equity** | **Focus Areas** |
|---|---|---|---|---|---|
| (Clark, Peterson et al. 2023) | Social Burden Metric | General | Tropical Cyclone, Extreme Temperature | C | Access |
| (Dong, Esmalian et al. 2020) | Integrated Physical-Social Vulnerability Assessment | Health | Tropical Cyclone, Flood | DSC | Access |
| (Logan and Guikema 2020) | Equitable Access to Essential Services | Food, Health, Power, Water, Sanitation, Communication | Tropical Cyclone | DSP | Access |
| (Esmalian, Dong et al. 2021) | Service Gap Model | Power | Tropical Cyclone | DSC | Capability |
| (Clark, Seager et al. 2018) | Capabilities Approach | Communication, Emergency, Power, Food, Health, Transportation, Water, Sanitation | General | C | Capability |
| (Abbou, Davidson et al. 2022) | Household Adaptations | Water, Power, Communication | General | DC | Capability |
| (Atallah, Djalali et al. 2018) | ABCD Roadmap | Health | General | DP | Community-centered approach |
| (Chen and Ji 2021) | Public Urgency Model | Power | Tropical Cyclone | S | Community-centered approach |
| (Yu and Welch 2022) | Agency Coordination Strategy | Transportation, Power, Communication, Stormwater | General | DSP | Community-centered approach |
| (Hsieh and Feng 2020) | DEMATEL | Transportation, Health | General | DSP | Community-centered approach |
| (Sweya, Wilkinson et al. 2021) | Social Resilience Tool | Water | Flood | DSPC | Community-centered approach |
| (Toland, Wein et al. 2023) | Community Vulnerability Assessment | Power, Water, Transportation, Food | Earthquake | DS | Vulnerability |
| (Coleman, Esmalian et al. 2020) | Anatomy of Susceptibility | Water, Transportation, Power, Health, Food, Sanitation | Tropical Cyclone Flood | DS | Vulnerability |
| (Daramola, Oni et al. 2016) | Pressure and Release Model | Water, Power, Health | General | DSC | Vulnerability |
| (Oswald Beiler and Mohammed 2016) | Transportation Justice Threshold Index Framework | Transportation | Flood | DS | Vulnerability |
| (Esmalian, Coleman et al. 2021) | Disruption Tolerance Index | Power, Transportation | Tropical Cyclone, Flood | DS | Vulnerability |
| (Esmalian, Wang et al. 2021) | Human-Hazard Nexus | Power | Tropical Cyclone, Flood | DC | Vulnerability |
| (Dhakal and Zhang 2023) | Social-Welfare-Based Infrastructure Resilience Assessment (SW-Infra-RA) | Power | Tropical Cyclone, Flood | D | Welfare economics |
| (Ulak, Yazici et al. 2020) | Prescriptive resilience model | Power | Tropical Cyclone | DP | Welfare economics |
| (Stock, Davidson et al. 2022) | Willingness to pay and unhappiness models | Water, Power | General | DC | Welfare economics |

It is notable to reflect on the intersections between the four dimensions of equity, see Figure 10. The most frequent link is between distributional-demographic and distributional-spatial (28), distributional-demographic and capacity (15), and distributional-demographic and procedural (6), which all revealed a connection to distributional-demographic equity.  There were comparatively fewer studies linking 3 dimensions including DSC (9), DCP (6), and DSP (6). Only 1 study had 4 connections. This can represent the current siloed approach to studying equity in infrastructure disrupted by hazard events.

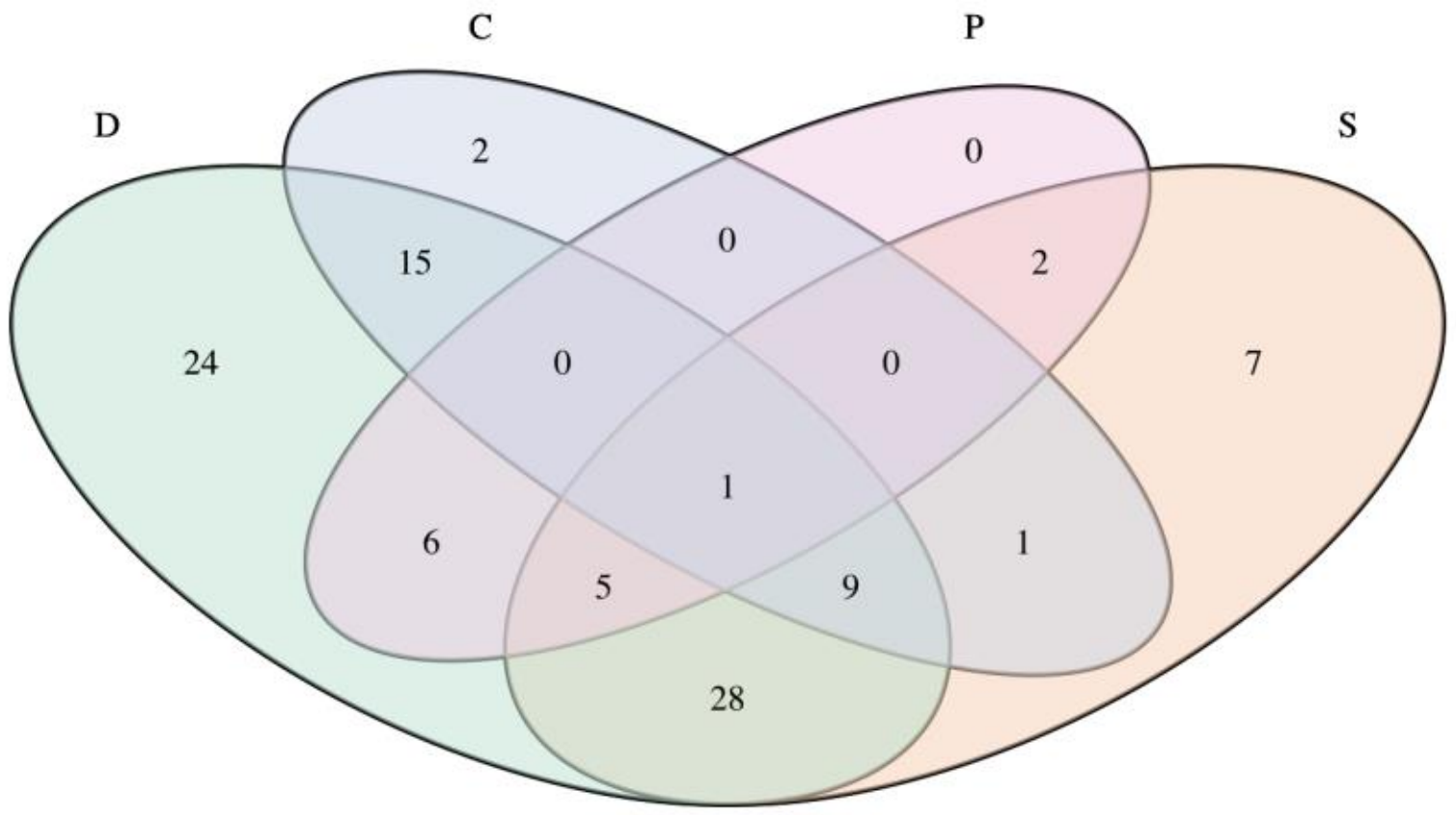

**Figure 10.** Venn-Diagram showing links between of DSPC dimensions of equity

Table 2 summarizes the primary findings of the systematic review of the equity literature, including what the studies are currently focusing on and the limitations in the overall view of equity in infrastructure management. We found a major geographic focus on the United States, and by extension on the global north while studies on the global south were limited. Not surprisingly then, cross-regional comparisons are also missing. This is in line with previous findings regarding resilience or the use of data-driven spatial approaches (Casali, Aydin et al., 2022). Regarding geographic scale, individual and local scales were the most prominent with a focus on open data and descriptive measures. For the temporal aspects, we confirm an increasing interest in the topic. We connected these to a disaster management cycle; however, the methods found in the literature could be clarified to provide clear temporal milestones in infrastructure recovery. Regarding data types and approaches, there was a dominance of open data that was analyzed with descriptive statistics and linear modeling approaches. Few approaches aimed to unravel the intricate relationships between equity, hazard and infrastructure with more complex modeling or analytical approaches such as Machine Learning or Agent-based modeling. Indeed, data limitations in infrastructure are still persistent (Calderon and Serven, 2014), and the "triangulation of data sources" could improve the understanding of infrastructure limitations (Hendricks et. al, 2018). Infrastructure systems like power, water, and health were most studied whereas stormwater, sanitation, and emergency were least studied. Correspondingly, water-related hazards (such as floods and hurricanes) dominate. Despite the recent rise in heatwaves, wildfires remain understudied in the equity realm. Among the equity dimensions, distributional equity received the most attention, and few studies considered procedural and capacity equity. The focus on minorities and lower-income households aligns with the other review findings of the general built environment (Seyedrezaei et. al, 2023). When focusing on the intersection of equity concept, hazard and infrastructure, we found that while there is a wide range of pathways, the majority of studies focused on flooding and hurricane events along with distributional-demographic and distributional-spatial equity aspects. Our review also identified a broad range of frameworks that stress different aspects of equity and relate to vulnerability, resilience, or welfare economics. What is missing though are analytical tools and approaches to integrate equity assessment into decision-making. Such findings parallel the work of Soden et. al. (2023) which found "only ~28 % attempt a quantitative evaluation of differential impacts in disasters".

The following section will elaborate on how to best address the research gaps identified in Table 2, and propose a vision statement for future research.

**Table 2.** Summary of the major focuses and limitation of equity literature

| **Dimension** | **Majority Focus** | **Research Gaps** |
|---|---|---|
| Geographic Location | • United States, and by extension the global north | • International Countries, particularly the global south |
| Geographic Scale | • Individual and local scales | • Cross regional and country comparisons |
| Temporal | • Increasing Trend of Publications<br>• Identified several timescales that integrate with disaster management cycle | • Vagueness in the methods of literature to how it applies to different timescales |
| Nature of the Study | ● Open-data and descriptive measures<br>● Descriptive statistics and linear regression | ● Location-intelligence and simulated datasets<br>● Machine learning (agent-based modeling, gravity network, generative modeling) |
| Hazard Event Type | ● Flooding and hurricane | ● Wildfire |
| Equity Dimension | ● Distributional-demographic and distributional-spatial | ● Procedural and Capacity |
| Infrastructure Type | ● Power, water, and transportation | ● Stormwater, emergency, and sanitation |
| Theoretical Perspective | ● Evidence for social and spatial inequities<br>● Conceptual measures of access | ● Analytical tools to disseminate equity<br>● Development of decision-making tools |

**Vision and Future Steps**

There is a need for a paradigm shift in how engineers, researchers, industry leaders, and communities are incorporating equity principles into infrastructure resilience. This review has shown that over the last decade, the primary research questions were to determine if there were inequities present in infrastructure disruptions during disasters. The resulting evidence from this collective research effort now shows a plethora of disparities captured through the experiences of different social groups, differential recovery patterns of spatial regions, the limiting capacities of households, and the inability to address vulnerabilities in institutional practices and policies. Disregarding the intentionality of infrastructure design and restoration, the outcomes strongly indicate that there are recurring disproportionate impacts in infrastructure loss events for vulnerable sub-populations. Now that the research has concluded the presence and surmounting extent of inequities of infrastructure management, we propose that the next era of research questions and objectives should be the (1) development of equity-centered decision-making tools, (2) weaving equity in computational models, and (3) monitoring equity performance with improved availability of quantitative data. Through principles of sustainability, accountability, and knowledge, such objectives would be guided by moving beyond distributional equity, recognizing understudied gaps of equity, and inclusion of all geographic regions, and by extension stakeholders (Figure 11).

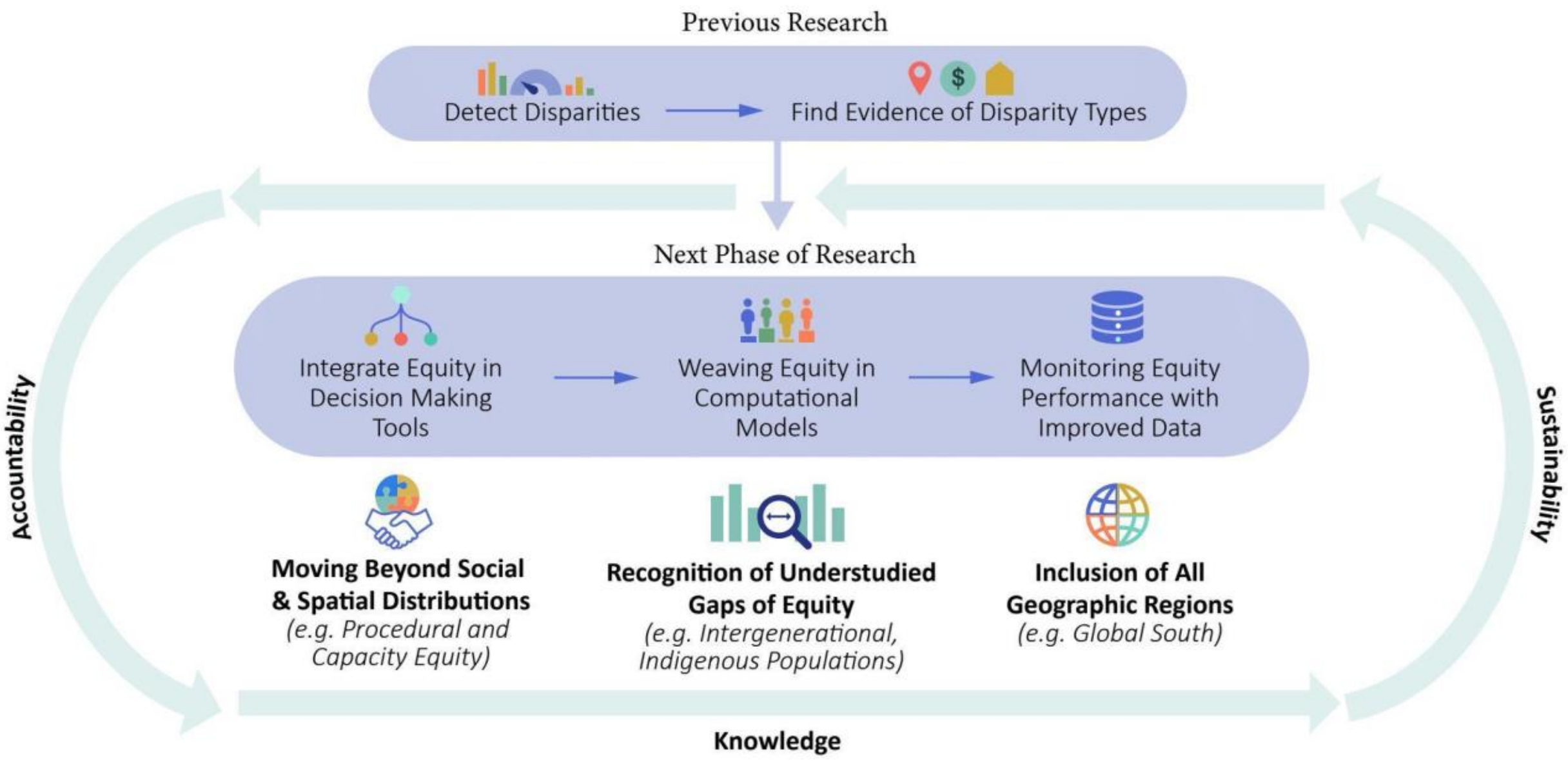

**Figure 11.** Vision for future research in equity of infrastructure management

**Integrating equity into decision-making tools:** Researchers have led preliminary efforts to characterize inequities of infrastructure management through conceptual frameworks and case studies. More efforts should be made to bridge the gap between the empirical understanding of equity to tangible tools for infrastructure planners, managers, engineers, and other decision-makers. It is not enough to discuss the importance of equity. Adequate key performance indicators and monitoring systems are needed to clarify the equity process. For instance, civil engineers often rely on the fundamentals of lifecycle and benefit-cost analysis to measure the success and failure of their infrastructure projects. Such fundamentals could be adapted in an equity mindset such as the proposed intergenerational discounting rate; however, it is important to recognize the flexibility of options for future generations (Teodoro et. al, 2023). Also, it is essential to quantify social costs of infrastructure service disruptions to properly determine the benefits of resilience investments. Current resilience investments often include significant upfront resource allocations and lack the proper consideration and quantification of social and environmental benefits of avoided infrastructure disruptions. Traditional standards of cost-benefit analyses used by infrastructure managers and operators, which solely focus on monetary gain, would not yield favorable results to support such significant investments to mitigate the human impacts of infrastructure losses. This limitation has been a hindering factor in infrastructure resilience because it contributes to delayed investments and inaction, resulting in unnecessary loss of services and social harm. Research is needed on the disparities in the deprivation costs of service losses for vulnerable populations. Hence, future research should strive to develop methods and frameworks to quantify social costs of infrastructure disruptions and integrate them into infrastructure resilience assessments, prioritization, and resource allocation. These tangible tools can ensure a standard baseline in approaching equity in infrastructure resilience processes which could then be tailored to the needs and characteristics of specific communities.

**Weaving equity in computational models:** In terms of analytical methods for infrastructure resilience assessments, most of the existing methods focus on physical networks and have paid limited attention to social dimensions and equity. Few studies (such as Esmalian et al. 2021) have created computational models to capture hazard-population-infrastructure interactions and quantify social impacts and disparities in addition to the evaluation of physical infrastructure performance. For analytical and computational methods to be able to inform equitable decision-making, future studies must integrate social dimensions of infrastructure resilience performance though the development of tools such as susceptibility curves (Esmalian et al. 2021) or empirical deprivation costs for infrastructure service losses.

**Monitoring equity performance with improved data:** Our literature review has also revealed the pitfalls to obtaining location intelligence and simulation data at more granular spatial scales to monitor infrastructure resilience performance and its equity. For instance, many studies have used reliable but ultimately approximate data sources for service outages, including human mobility, satellite, and telemetry-based data. Thus, we encourage the shift towards transparent and open datasets from utility companies in service disruption and outage events. This transparency fosters an environment of accountability and innovation to ensure that equity standards are being diligently applied in infrastructure management. The greater availability of data also provides researchers with an accurate understanding of infrastructure losses and disparities across different subpopulations which will potentially cultivate evidence-based recommendations for future events. Also, the availability of such data will enable monitoring of the extent to which hazard mitigation and infrastructure restoration practices of infrastructure agencies have been equitable.

**Broader geographical contexts:** The studies on equity in infrastructure resilience have primarily focused on the global north. Very few studies have examined equitable infrastructure resilience practices and methods for the global south, which outlines the different infrastructure challenges (e.g., intermittent services) and data availability in the region. In fact, the dearth of studies on equitable infrastructure resilience in the Global South could contribute to greater inequality in those regions due to the absence of empirical evidence and proper methodological solutions. This aligns with other findings on sustainable development goals and climate adaptation broadly (Berrang-Ford et al., 2021). Global research efforts that include international collaborations among researchers across the global north and south regions can bridge this gap and expand the breadth of knowledge and solutions for equitable infrastructure resilience.

**Recognition of Understudied Gaps of Equity:** Even though intergenerational justice issues have increasingly sparked attention on the climate change discussion, intergenerational equity issues in infrastructure resilience assessments have received limited attention. We argue that intergenerational equity warrants special attention as infrastructure systems have long life cycles that span across multiple generations, and ultimately the decisions on financing, restoration, and new construction will have a significant impact on the ability of future generations to withstand the impact of stronger climate hazard events. Non-action may lead to tremendous costs in the long run (Teodoro et al., 2023). It is the responsibility of current research to understand the long-term effects of equity in infrastructure management to mitigate future losses and maintain the flexibility of future generations. As a means of procedural justice, these generations should have the space to make choices, instead of being locked in by the decisions that we make today. Hence, future studies should develop methods to measure and integrate intergenerational inequity in infrastructure resilience assessments. In addition, certain demographic groups such as indigenous populations and persons with disabilities have not been sufficiently studied. Indigenous populations face significant geographical, cultural, and linguistic barriers that make their experiences with disrupted infrastructure services distinct from those of the broader population. Similarly, persons with disabilities require specific attention to access critical services. As a few examples, these knowledge gaps hinder our ability to develop equitable disaster strategies that address the unique needs and vulnerabilities of these underrepresented communities during times of infrastructure disruption.

**Moving beyond distributional and spatial equity:** Finally, while significant attention has been paid to distributional and spatial inequity issues in infrastructure disruptions, there remain several understudied aspects of equity. Even though the procedural and capacity dimensions of equity hold the greatest potential for people to feel more included in the infrastructure resilience process, these dimensions are not as discussed in research studies. Instead of depending directly on the infrastructure systems, individual households can adapt to disrupted periods through substituted services and alternative actions (such as Abbou, Davidson et al. 2022). Citizen-science research or participatory studies can begin by empowering locals to understand and monitor their resilience (Champlin et al., 2023) or failures in their infrastructure systems (such as Gharaibeh, Oti et al. 2021). Future studies should expand inquiries regarding the procedural and capacity dimension of equity in infrastructure resilience assessments and management.

**Acknowledgements:** This material is based in part upon work supported by the National Science Foundation under Grant CMMI-1846069 (CAREER) and the support of the National Science Foundation Graduate Research Fellowship. We would like to thank the contributions of our undergraduate students: Nhat Bui, Shweta Kumaran, Colton Singh, and Samuel Baez.

**Author Contributions**: All authors critically revised the manuscript, gave final approval for publication, and agree to be held accountable for the work performed therein. N.C. was the lead Ph.D. student researcher and first author, who was responsible for guiding data collection, performing the main part of the analysis, interpreting the significant results, and writing most of the manuscript. X.L was responsible for guiding data collection, figure creations, and assisting in manuscript. T.C and A. M were the faculty advisors for the project and provided critical feedback on the literature review development, analysis and manuscript.

**Data Availability:** The created excel database which includes information on the key parts of the 8-dimensional equity framework will be uploaded to DesignSafe-CI.

**Supplemental Information**

The following figures provide additional context to the findings.

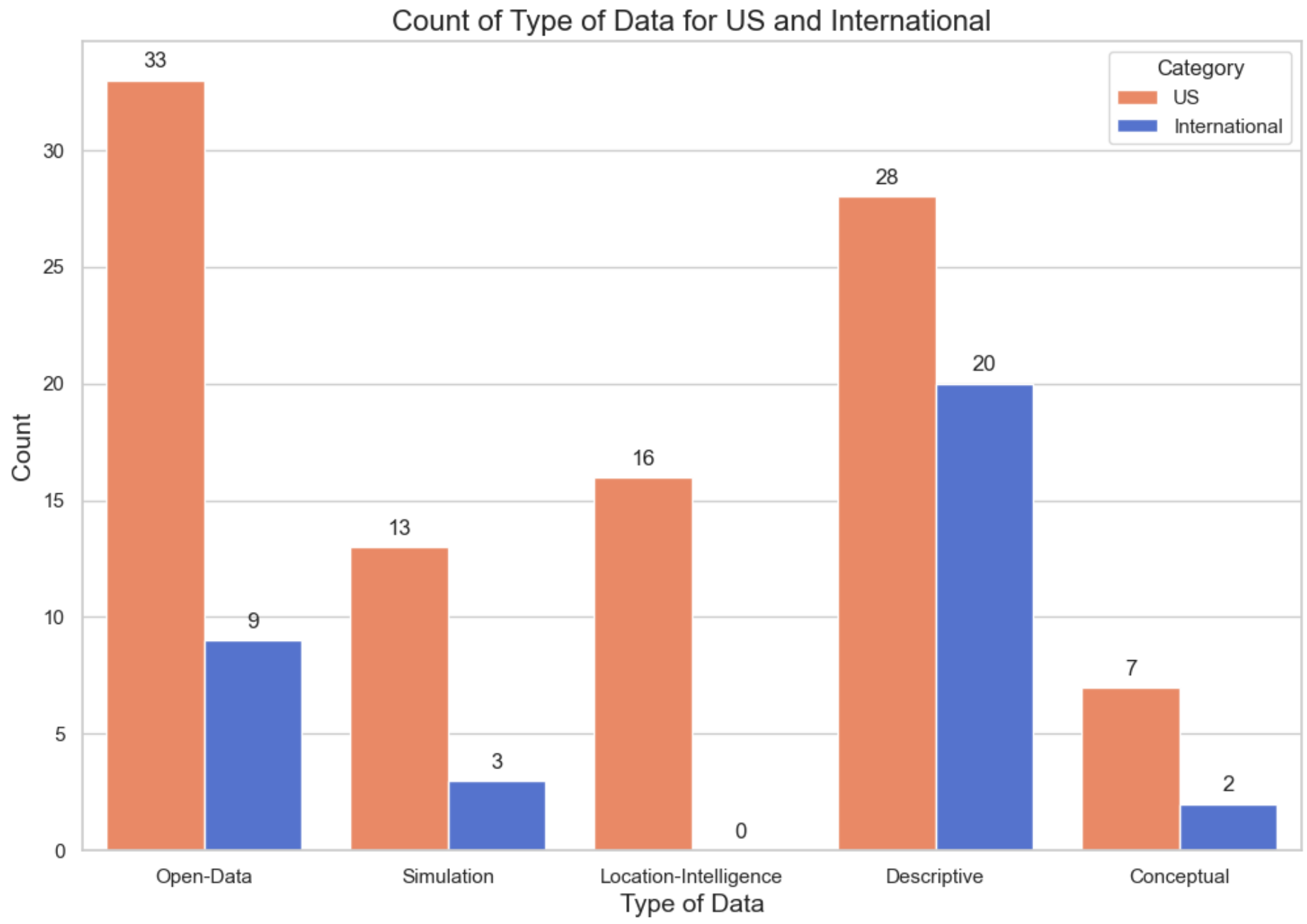

**Figure 1A.** Frequency of Data Type between US and International studies. As shown, there are no location-intelligence studies for international studies.

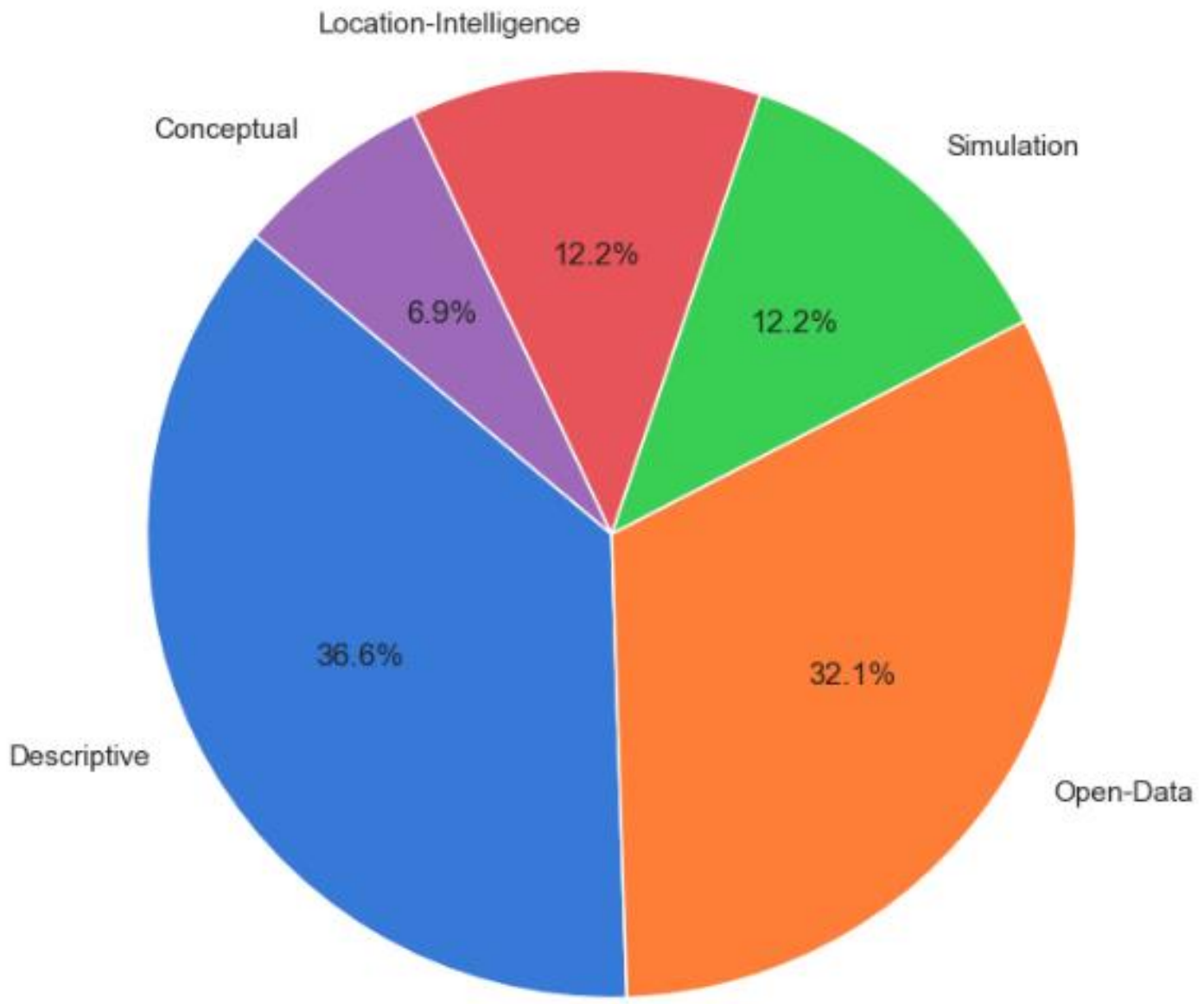

**Figure 2A.** Percentage of Data Type of All Studies. In total, open-data and descriptive data is most commonly used, followed by simulation and location-intelligence, and lastly conceptual.

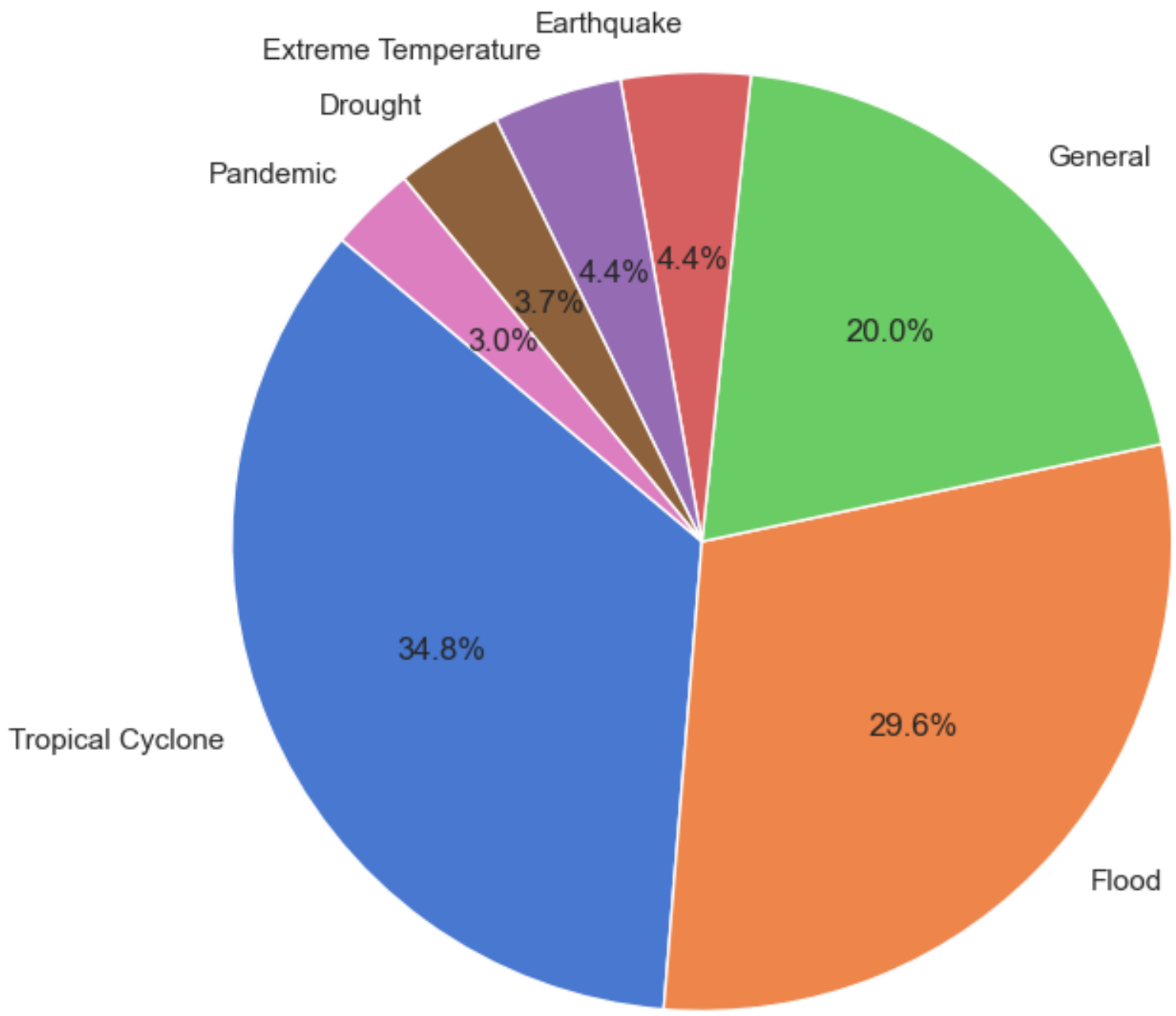

**Figure 4A.** Percentage of Hazards of All Studies. Flood and tropical cyclone make up over half the studied hazards.

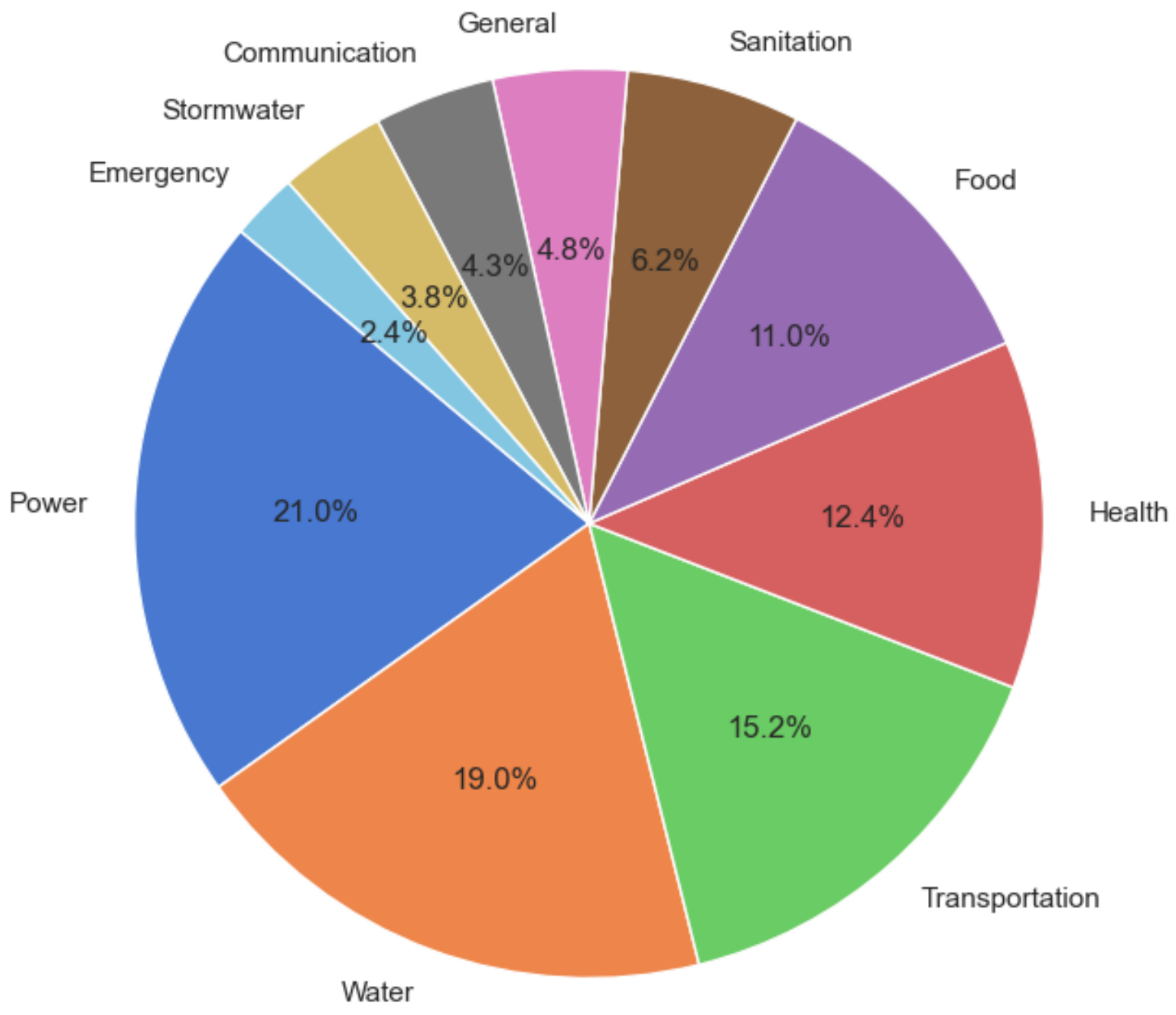

**Figure 5A.** Percentage of Transportation Systems of All Studies. Power, water, and transportation are the top studied.

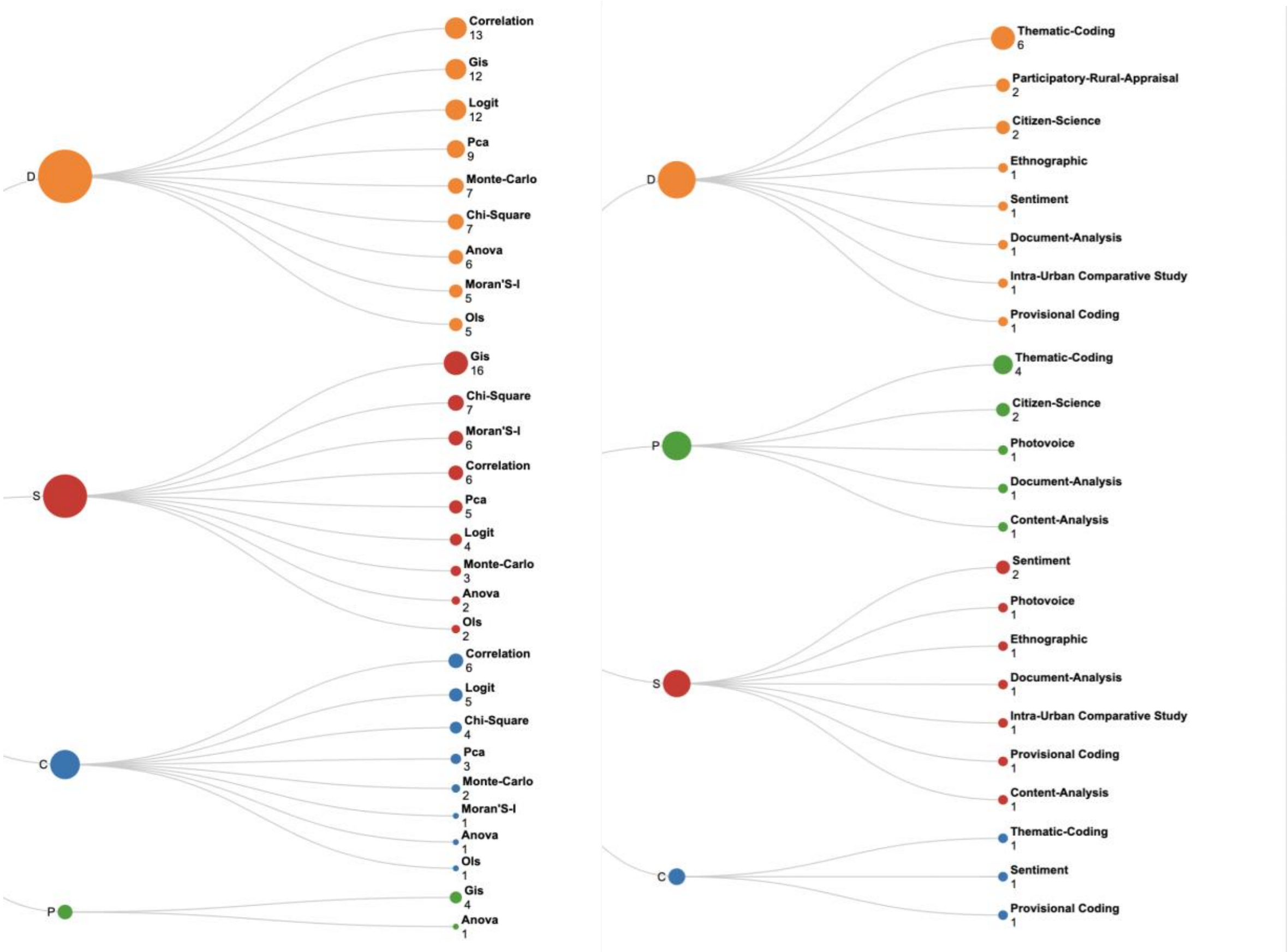

**Figure 6A.** Flow diagram to reveal the different types of quantitative (left) and qualitative (right) methods

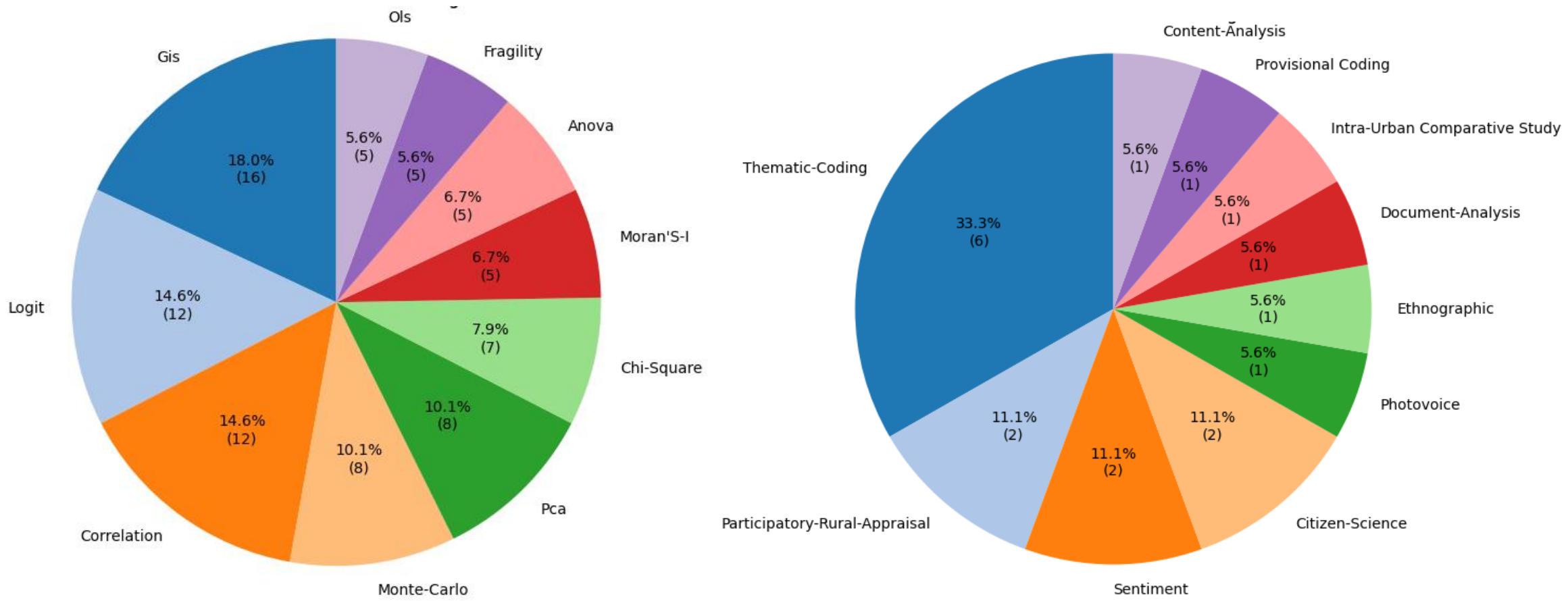

**Figure 7A.** Percentage of the top ten types of quantitative (left) and qualitative (right) methods

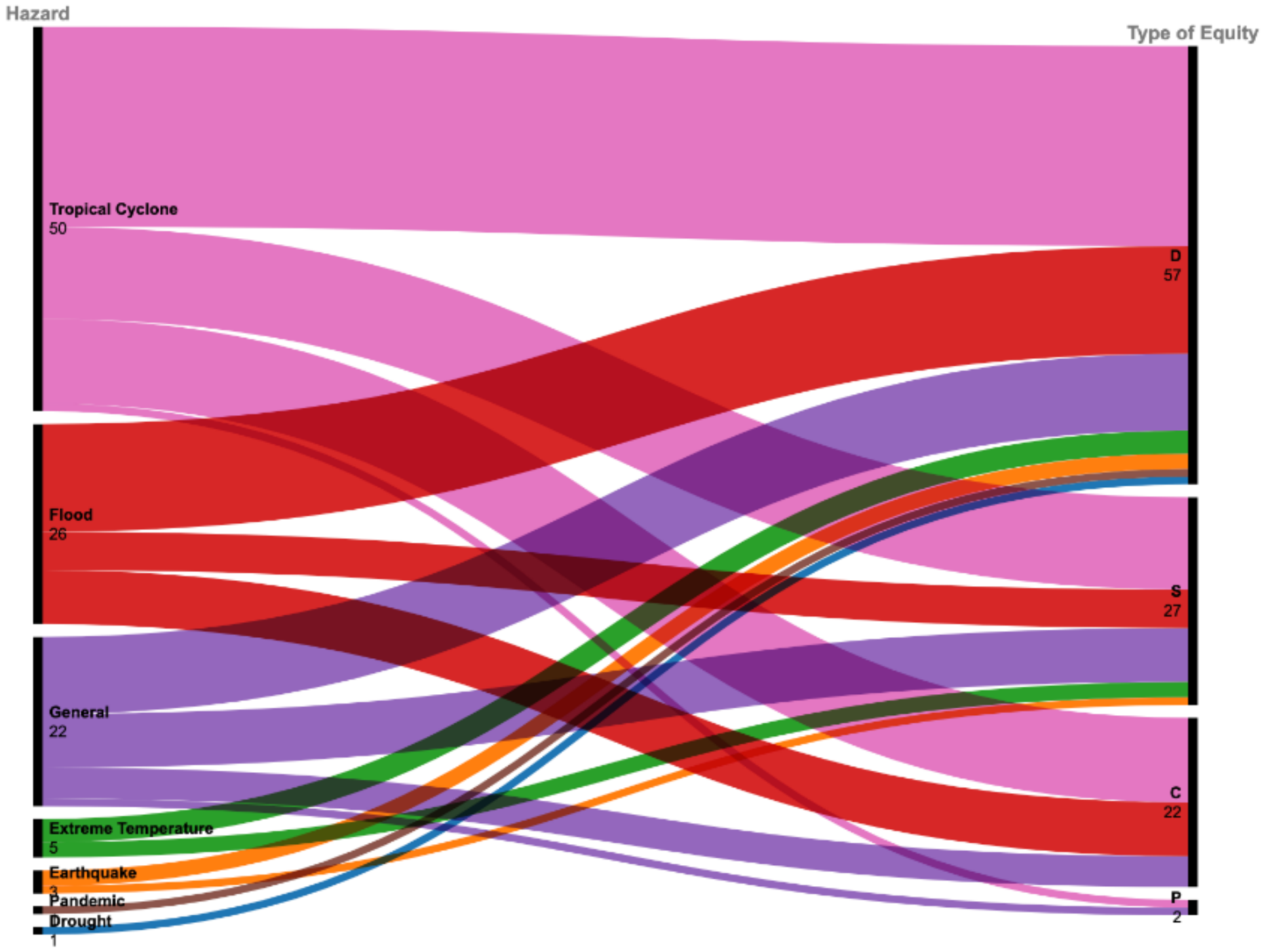

**Figure 8A.** Sankey diagram of power disruptions to types of hazard and equity dimensions

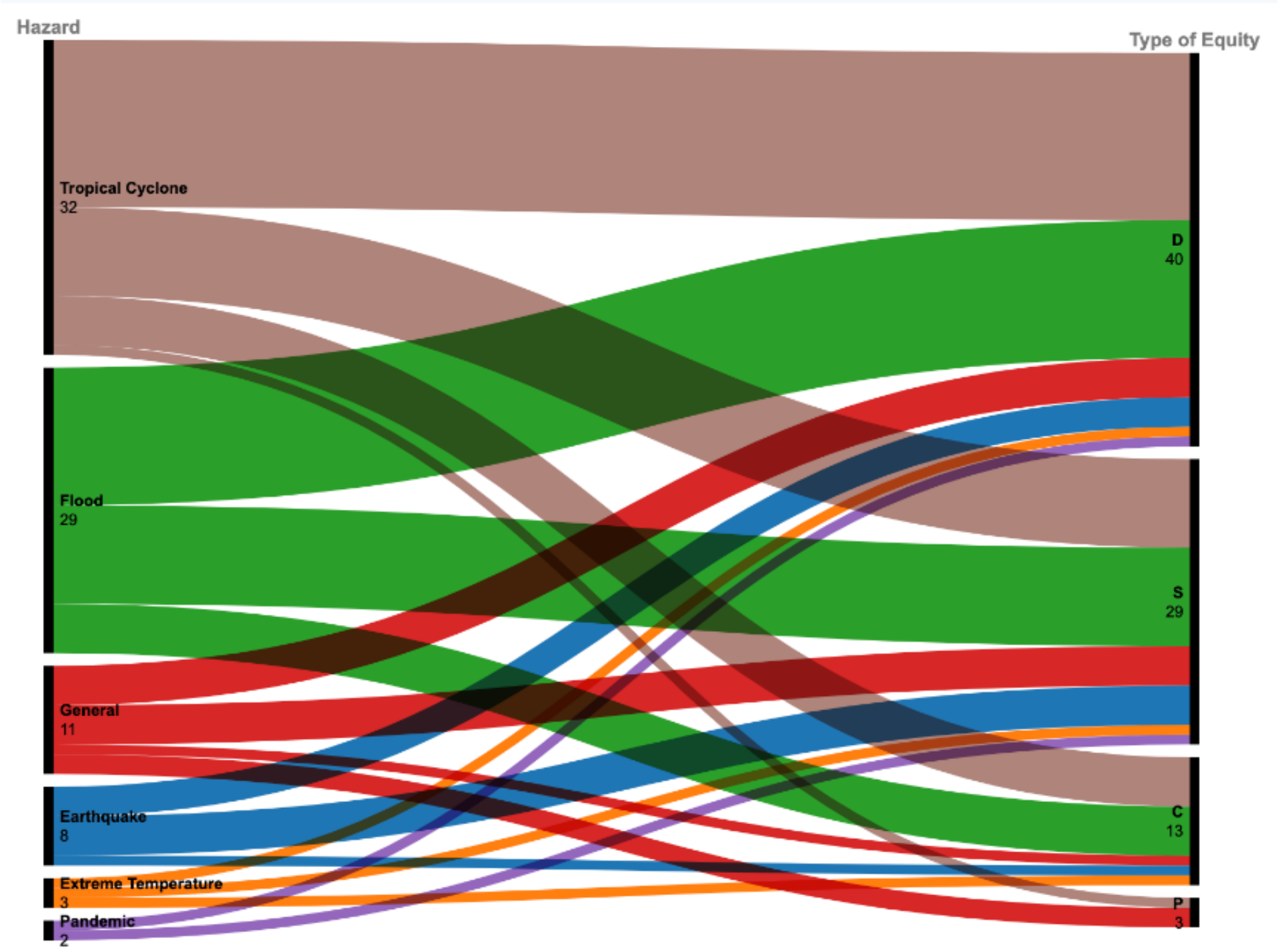

**Figure 9A.** Sankey diagram of transportation disruptions to types of hazard and equity dimensions

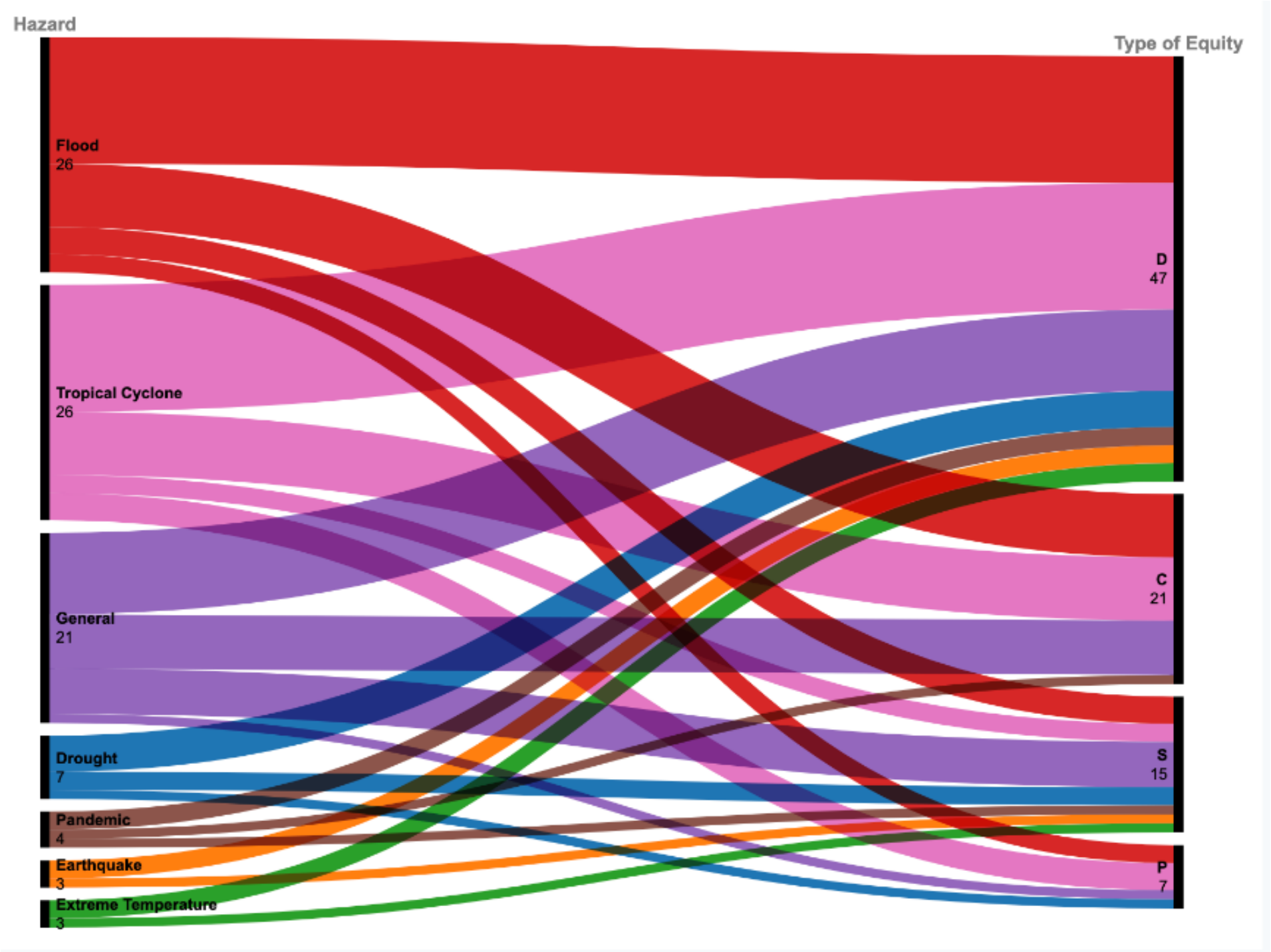

**Figure 10A.** Sankey diagram of water disruptions to types of hazard and equity dimensions

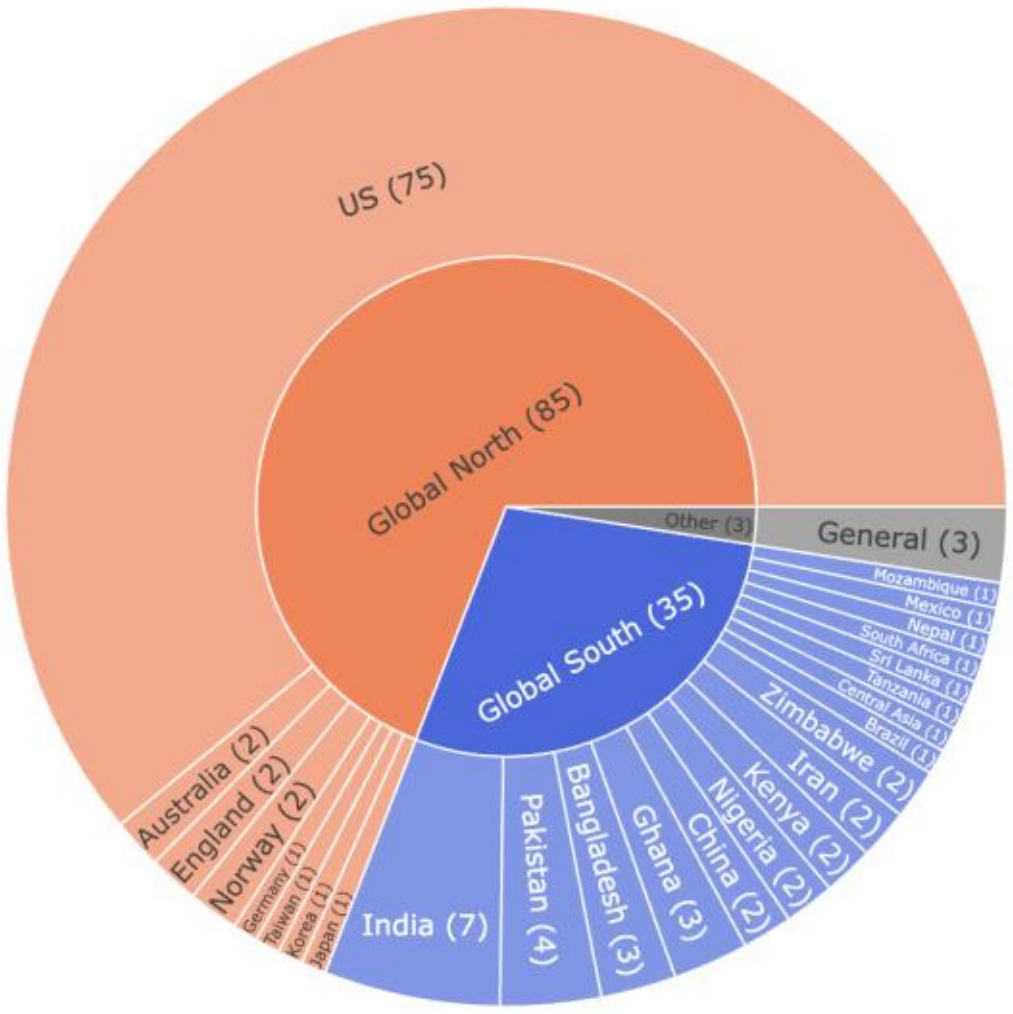

**Figure 11A**. Distribution of global north and global south by country

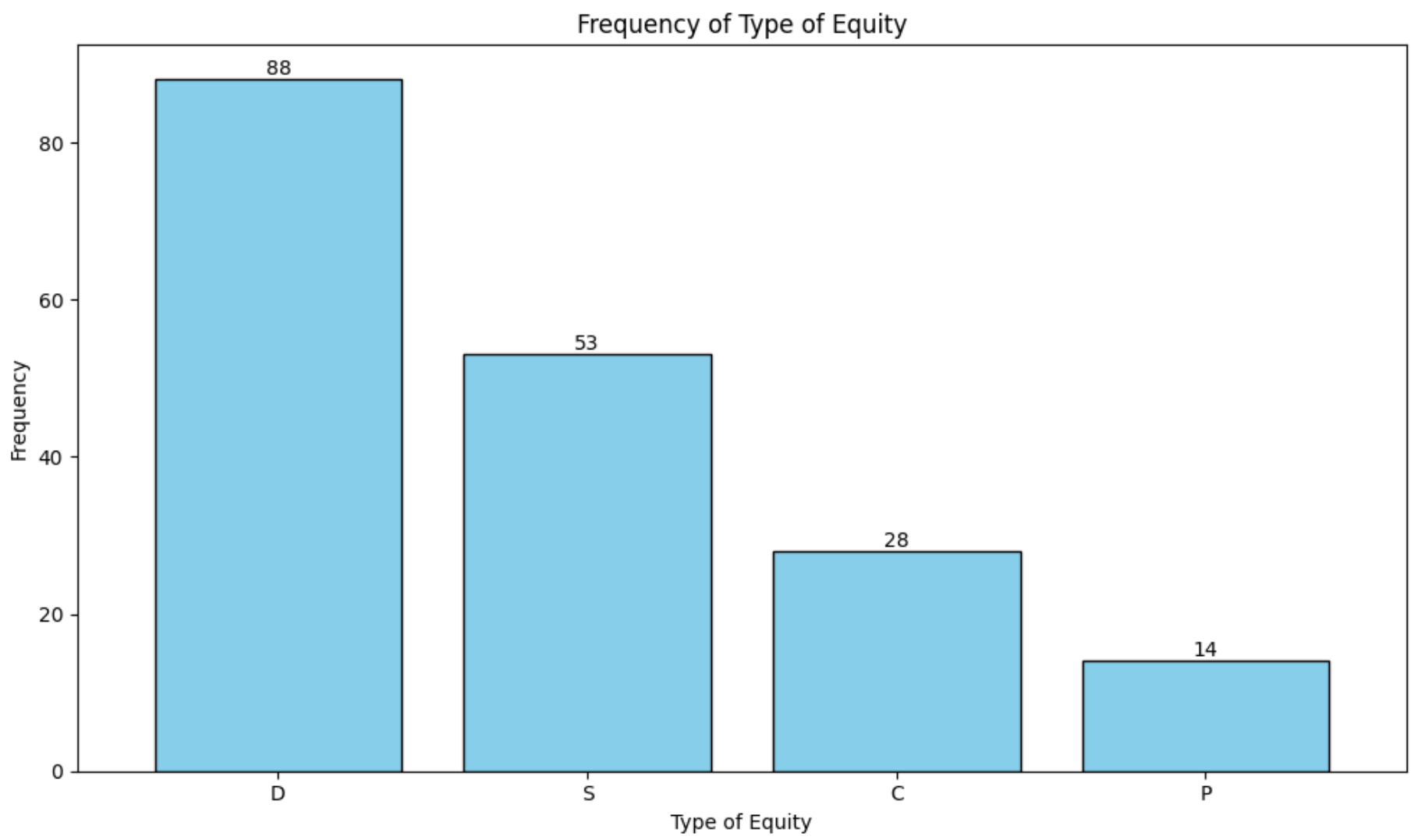

**Figure 12A.** Frequency of equity dimensions